\documentclass[twocolumn,tighten,twocolappendix]{aastex631}
\shorttitle{GENERALIZED CORONAL LOOP SCALING LAWS}
\shortauthors{Dai et al.}
\usepackage{amsmath}
\usepackage{apjfonts}

\begin{document}
\title{Generalized Coronal Loop Scaling Laws and Their Implication for Turbulence in Solar Active Region Loops}
		
\author{Y.~Dai}
\affiliation{School of Astronomy and Space Science, Nanjing University, Nanjing 210023, People's Republic of China}
\affiliation{Key Laboratory of Modern Astronomy and Astrophysics (Nanjing University), Ministry of Education, Nanjing 210023, People's Republic of China}

\author{J.~J.~Xiang}
\affiliation{School of Astronomy and Space Science, Nanjing University, Nanjing 210023, People's Republic of China}
		
\author{M.~D.~Ding}
\affiliation{School of Astronomy and Space Science, Nanjing University, Nanjing 210023, People's Republic of China}
\affiliation{Key Laboratory of Modern Astronomy and Astrophysics (Nanjing University), Ministry of Education, Nanjing 210023, People's Republic of China}

\correspondingauthor{Yu~Dai}
\email{ydai@nju.edu.cn}
		
\begin{abstract}
Recent coronal loop modeling has emphasized the importance of combining both Coulomb collisions and turbulent scattering to characterize field-aligned thermal conduction, which invokes a hybrid loop model. In this work we generalize the hybrid model by incorporating nonuniform heating and cross section that are both formulated by a power-law function of temperature. Based on the hybrid model solutions, we construct scaling laws that relate loop-top temperature ($T_a$) and heating rate ($H_a$) to other loop parameters. It is found that the loop-top properties for turbulent loops are additionally power-law functions of turbulent mean free path ($\lambda_T$), with the functional forms varying from situation to situation that depends on the specification of the heating and/or areal parameters. More importantly, both a sufficiently footpoint-concentrated heating and a cross-sectional expansion with height can effectively weaken (strengthen) the negative (positive) power-law dependence of $T_a$ ($H_a$) on $\lambda_T$. The reason lies in a notable reduction of heat flux by footpoint heating and/or cross-sectional expansion in the turbulence-dominated coronal part, where turbulent scattering introduces a much weaker dependence of the conduction coefficient on temperature. In this region, therefore, the reduction of the heat flux predominately relies on a backward flattening of the temperature gradient. Through numerical modeling that incorporates more realistic conditions, this scenario is further consolidated. Our results have important implication for solar active region (AR) loops. With the factors of nonuniform heating and cross section taken into account, AR loops can bear relatively stronger turbulence while still keeping a physically reasonable temperature for nonflaring loops.
\end{abstract}
		
\keywords{Solar active regions (1974), Solar corona (1483), Solar coronal loops (1485), Hydrodynamics (1963)}
	
\section{introduction} 
It has been revealed by the extreme-ultraviolet (EUV) and/or X-ray imaging observations from a variety of space-borne missions \citep[e.g.,][]{Kosugi07,Kaiser08,Pasnell12,Muller20} that loop structures constitute the bulk of magnetically closed solar corona. In a low-$\beta$ coronal atmosphere, magnetic fields not only determine the loop geometry, but also harbor energy that powers the loop heating. Depending on the concentration and connectivity of loop-hosting magnetic fields, coronal loops appear either as diffuse background over the quiet Sun, or like discrete arcade-shaped features above active regions (ARs). Appropriate modeling of these loops provides crucial information on the configuration of coronal magnetic fields and the nature of coronal heating \citep{Martens10,Reale14,Klimchuk20}.

Except for few transient brightening periods, nonflaring coronal loops typically appear quasi-static or are just slowly evolving over most of their lifetime. To lowest order, this leads to hydrostatic modeling of the loops. In a pioneering work, \citet{Rosner78} studied the energy balance between optically thin radiative losses, volumetric heating, and field-aligned thermal conduction in quasi-static loops, based on which they established the well-known ``Rosner--Tucker--Vaiana (RTV)" scaling laws that relate the maximum temperature (typically located at the loop apex) and heating rate of a loop to its pressure and half-length. The RTV scaling laws are of immense value, for their comparisons with observations can help us test the validity of simple static modeling or otherwise incorporate new physics in more complex modeling to reconcile with the observations. 

In deriving the RTV scaling laws, it was assumed that the conductive transfer of heat flux is controlled by collisional scattering of thermal electrons. With such form of thermal conduction, the model predicts a differential emission measure (DEM) profile that decreases from coronal temperatures and maintains a constant level through transition region (TR) temperatures \citep{Rosner78}. The DEM profile is broadly consistent with observations at coronal temperatures, but fails to reproduce the observed DEMs toward lower TR ($<10^5$~K), where a large excess to the coronal level is revealed \citep[e.g.,][]{Raymond81,delZanna18}. The discrepancy could be reconciled by involving external structures such as cool low-lying loops and/or spicules \citep[e.g.,][]{Antiochos86,dePontieu11}. One can also incorporate new physics to affect the transport of heat flux inside the loop, and consequently the shape of DEM, both of which are directly related to temperature gradient. Turbulent scattering, which arises from magnetic field fluctuations within the loop, serves as a candidate mechanism \citep{Bian16}.

In explosive phenomena such as solar flares, strong turbulence is self-consistently generated in reconnection current sheets as well as in post-flare loops \citep[e.g.,][]{Cheng18,Wang23}. The turbulent scattering gives rise to a variety of observational signatures, including nonthermal broadening of flare emission lines \citep[e.g.,][]{Young15,Li18}, prominent coronal sources observed in hard X-ray bands \citep[e.g.,][]{Masuda94,Dai10,Guo12}. Due to the magnetic concentration in ARs, it is believed that a certain level of turbulence is also present in nonflaring AR loops. \citet{Emslie22} have recently developed a hybrid model for AR loops, where Coulomb collisions and turbulent scattering dominate the thermal conduction in different regimes of a loop. Based on the model, temperature and DEM profiles are calculated and scaling laws are derived, both of which show substantial changes from the wholly collisional case. Owing to the suppression of conduction coefficient by turbulent scattering, the DEM with the hybrid model exhibits an increase toward low temperatures, more compatible with observations. Nevertheless, the inclusion of turbulence also significantly elevates the loop temperature (even to an unrealistically high level), which seems to impose a strong constraint on the turbulence strength in AR loops.

Both the original wholly collisional model of \citet{Rosner78} and the recent hybrid model of \citet{Emslie22} assume a uniform background heating and loop cross section. In fact, the decay of coronal magnetic fields with height implies that neither the volumetric heating nor the cross section can keep constant along a loop. Such factors have been considered in several loop modeling works \citep[e.g.,][]{Vesecky79,Serio81,Winebarger03,Martens10,Cargill22}. In the fame of collision-dominated thermal conduction throughout the loop, the results just show marginal modifications to the original RTV scaling laws. When turbulent scattering is further included, nevertheless, how will nonuniform heating and cross section affect the loop characteristics in terms of the scaling laws? In this work, we generalize the hybrid model of \citet{Emslie22} by incorporating both nonuniform heating and variable loop cross section, and construct scaling laws for various situations. It is found that the functional forms of the scaling laws depend on the specification of the heating and/or loop areal parameters, and both a sufficiently footpoint-concentrated heating and a cross-sectional expansion with height can effectively weaken the dependence of loop-top temperature on turbulence strength (quantified by turbulent mean free path). The results have important implication for AR loops, in which the former strong constraint on  turbulence strength in terms of loop temperature can be greatly relaxed.

The rest of the paper is organized as follows. We describe the generalized hybrid model in Section \ref{sec2} and deduce our scaling laws in Section \ref{sec3}. The effects of nonuniform heating and cross section are explored based on the hybrid model solutions in Section \ref{sec4} and numerically validated in Section \ref{sec5}. Finally, we discuss the results and draw our conclusions in Section \ref{sec6}.

\section{Generalized Hybrid Loop Model}\label{sec2}
We generalize the hybrid loop model of \citet{Emslie22} by incorporating both nonuniform heating and variable loop cross section. For a coronal loop in hydrostatic equilibrium, the equation of energy conservation is expressed as
\begin{equation}
\frac{1}{A}\frac{d}{ds}\left(AF_{\scriptscriptstyle\! C}\right)=E_{\scriptscriptstyle\! R}+E_{\scriptscriptstyle\! H},
\end{equation}
where $s$ is the distance along the loop measured upward from the loop base, $A$ is the loop cross-sectional area, $F_{\scriptscriptstyle\! C}=-\kappa dT/ds$ is the field-aligned heat flux with $T$ being the loop temperature and  $\kappa$ representing a model-dependent conduction coefficient,  $E_{\scriptscriptstyle\! R}$ is the radiative losses from the loop, and $E_{\scriptscriptstyle\! H}$ is the volumetric loop heating. Here we assume a semi-circular loop with a full length of $2L$, and due to the symmetry, we  consider only half of the loop, along which the temperature monotonically increases from $T_b$ at either footpoint of the loop ($s=0$) to $T_a$ at the loop apex ($s=L$).

The coronal radiative losses are believed to be optically thin, and hence are formulated as
\begin{equation}
E_{\scriptscriptstyle\! R}=-n_e^2\Phi(T),
\end{equation}
where $n_e$ is the electron number density and $\Phi(T)$ is an optically thin radiative loss function. In analytical modeling the radiation function is typically simplified as a single power-law form of $\Phi(T)=\chi_0T^{-\gamma}$, with $\chi_0$ being the proportionality coefficient and $\gamma$ ($\gamma>0$)  the slope. Further applying the equation of state for fully ionized plasma $P=2n_ek_{\scriptscriptstyle\! B}T$ (where $P$ denotes the gas pressure assumed to be uniform throughout the loop for analytical convenience, and $k_{\scriptscriptstyle\! B}=1.38\times10^{-16}$ erg K$^{-1}$  is the Boltzmann constant), the radiative loss term is accordingly rewritten as
\begin{equation}
E_{\scriptscriptstyle\! R}=-\frac{\chi_0P^2}{4k_{\scriptscriptstyle\! B}^2}T^{-(2+\gamma)}.
\end{equation}

As formulated in \citet{Martens10},  the loop heating rate and cross-sectional area are also assumed to follow a power-law dependence on the temperature, i.e., 
\begin{equation}
E_{\scriptscriptstyle\! H}=H_a\left(\frac{T}{T_a}\right)^{\alpha}, \ \   A=A_a\left(\frac{T}{T_a}\right)^{\delta},
\end{equation}
 where $H_a$ and  $A_a$ are the corresponding properties at the loop apex, and indices $\alpha$ and $\delta$ ($\delta\ge0$) quantify the degrees of heating stratification and loop cross-sectional expansion, respectively. Note that the particular case of $\alpha=0$ (uniform heating) and $\delta=0$ (constant loop cross section) has been modeled in \citet{Emslie22}.
 
Depending on the environmental conditions, the conductive transfer of the heat flux can be controlled by various physical processes. In each process, a pertinent mean free path $\lambda$ is related to the conduction coefficient by
\begin{equation}
\kappa=2n_ek_{\scriptscriptstyle\! B}v_{\mathrm{th}}\lambda,\label{eqnkappa}
\end{equation}
where $v_{\mathrm{th}}=(2k_{\scriptscriptstyle\! B}T/m_e)^{1/2}$ is the thermal velocity of electrons with $m_e=9.1\times10^{-28}$ g being the electron mass. For thermal conduction dominated by Coulomb collisions,  the mean free path is modeled with the Spitzer--H\"{a}rm approximation \citep{Spitzer53,Spitzer62}, which leads to a collisional mean free path $\lambda_C$ of
\begin{equation}
\lambda_C=\frac{(2k_{\scriptscriptstyle\! B}T)^2}{2\pi e^4n_e\ln\Lambda}=\frac{c_{\scriptscriptstyle\! R}T^3}{P},\label{eqnlambc}
\end{equation}
where $e=4.8\times10^{-10}$ esu is the electron charge and $\ln\Lambda\approx20$ is the Coulomb logarithm. For simplicity we make the identification $c_{\scriptscriptstyle\! R}=4k_{\scriptscriptstyle\! B}^3/\pi e^4 \ln\Lambda\approx3.15\times10^{-12}$ erg cm$^{-2}$ K$^{-3}$ as used in \citet{Bradshaw19}. Substituting the form of $\lambda_C$ into Equation (\ref{eqnkappa}), it yields
\begin{equation}
\kappa=\left(\frac{2k_{\scriptscriptstyle\! B}}{m_e}\right)^{1/2}c_{\scriptscriptstyle\! R}T^{5/2}=\kappa_{\scriptscriptstyle 0C}T^{5/2},\label{kappacdc}
\end{equation}
where $\kappa_{\scriptscriptstyle 0C}=({2k_{\scriptscriptstyle\! B}}/{m_e})^{1/2}c_{\scriptscriptstyle\! R}=1.73\times10^{-6}$ erg cm$^{-1}$ s$^{-1}$ K$^{-7/2}$ is the classical Spitzer conductivity for collision-dominated conduction (CDC).

In case of strong turbulence, the conductive heat transport would be alternatively controlled by turbulent scattering \citep{Bian16,Emslie18}. Following the simplification made in \citet{Bian18,Bradshaw19}, we assume a constant turbulent mean free path $\lambda_T$ that is independent on the loop temperature and density. Then the conduction coefficient for turbulence-dominated conduction (TDC) is formulated as
\begin{equation}
\kappa=\left(\frac{2k_{\scriptscriptstyle\! B}}{m_e}\right)^{1/2}\lambda_TPT^{-1/2}=\kappa_{\scriptscriptstyle 0T}T^{-1/2},\label{kappatdc}
\end{equation}
where we make the identification $\kappa_{\scriptscriptstyle 0T}=({2k_{\scriptscriptstyle\! B}}/{m_e})^{1/2}\lambda_TP$ for simplicity. Note that $\kappa_{\scriptscriptstyle 0T}$ is also a constant as we assume a constant loop pressure $P$ and turbulent mean free path $\lambda_T$ throughout the loop.

With the above two processes, the hybrid model combines both a collision-dominated domain and a turbulence-dominated domain within a single loop. In the lower part of the loop where the collisional mean free path is short owing to small values of the temperature, the thermal conduction is dominated by collisions, with the conduction coefficient being determined by Equation (\ref{kappacdc}). As the collisional mean free path rapidly increases with temperature and finally exceeds the turbulent mean free path, turbulent scattering will take over in the upper part, and hence Equation (\ref{kappatdc}) applies.

We first introduce the dimensionless temperature and coordinate
\begin{equation}
\theta=\frac{T}{T_a}, \ \  x=\frac{s}{L}.
\end{equation}
For the collision-dominated part, we define an auxiliary variable $\eta$ that is related to the temperature by
\begin{equation}
\eta=\theta^{(7+2\delta)/2},
\end{equation}
and two dimensionless parameters
\begin{equation}
\varepsilon_1=\frac{8k_{\scriptscriptstyle\! B}^2\kappa_{\scriptscriptstyle 0C}T_a^{(11+2\gamma)/2}}{(7+2\delta)\chi_0P^2L^2}, \ \ \xi=\frac{4k_{\scriptscriptstyle\! B}^2H_aT_a^{2+\gamma}}{\chi_0P^2}.\label{epsilonxi}
\end{equation}
Then the energy equation for the collision-dominated part is cast in a dimensionless form of 
\begin{equation}
\varepsilon_1\frac{d^2\eta}{dx^2}=\eta^{\mu_1}-\xi\eta^{\nu_1},\label{eecoll}
\end{equation}
where
\begin{equation}
\mu_1=-\frac{2(2+\gamma-\delta)}{7+2\delta},\ \ \nu_1=\frac{2(\alpha+\delta)}{7+2\delta}.
\end{equation}

Multiplying with $d\eta/dx$ on both sides, and applying the boundary conditions of $\eta=\eta_b=\theta_b^{(7+2\delta)/2}=(T_b/T_a)^{(7+2\delta)/2}$ and $d\eta/dx=0$ (vanishing heat flux) at $x=0$, the integration of Equation (\ref{eecoll}) results in a first integral of
\begin{equation}
\frac{\varepsilon_1}{2}\left(\frac{d\eta}{dx}\right)^2=\frac{\eta^{\mu_1+1}-\eta_b^{\mu_1+1}}{\mu_1+1}-\frac{\xi(\eta^{\nu_1+1}-\eta_b^{\nu_1+1})}{\nu_1+1}.\label{ficoll}
\end{equation}

On the other hand, by introducing another temperature-related variable 
\begin{equation}
\zeta=\theta^{(1+2\delta)/2},
\end{equation}
and an additional dimensionless parameter
\begin{equation}
\varepsilon_2=\frac{8k_{\scriptscriptstyle\! B}^2\kappa_{\scriptscriptstyle 0T}T_a^{(5+2\gamma)/2}}{(1+2\delta)\chi_0P^2L^2},\label{epsilon2}
\end{equation}
the dimensionless energy equation for the turbulence-dominated part is similarly cast to
\begin{equation}
\varepsilon_2\frac{d^2\zeta}{dx^2}=\zeta^{\mu_2}-\xi\zeta^{\nu_2},\label{eeturb}
\end{equation}
where
\begin{equation}
\mu_2=-\frac{2(2+\gamma-\delta)}{1+2\delta},\ \ \nu_2=\frac{2(\alpha+\delta)}{1+2\delta}.
\end{equation}

Using the boundary conditions of $\zeta=1$ and $d\zeta/dx=0$ (local temperature maximum) at $x=1$, the integration of Equation (\ref{eeturb})  yields another first integral of
\begin{equation}
-\frac{\varepsilon_2}{2}\left(\frac{d\zeta}{dx}\right)^2=\frac{1-\zeta^{\mu_2+1}}{\mu_2+1}-\frac{\xi(1-\zeta^{\nu_2+1})}{\nu_2+1}.\label{fiturb}
\end{equation}

The interface between the collision-dominated domain and the turbulence-dominated domain lies at a dimensionless coordinate of $\ell$, at which the temperature is $T_0$ (or $\theta_0=T_0/T_a$ in dimensionless form, and correspondingly converted to $\eta_0$ and $\zeta_0$). By this definition, it is evident that
\begin{equation}
\lambda_T=\lambda_C(T_0)=\frac{c_{\scriptscriptstyle\! R}T_0^3}{P}=\frac{c_{\scriptscriptstyle\! R}T_a^3}{P}\theta_0^3\label{lambt},
\end{equation}
which further leads to
\begin{equation}
\frac{\kappa_{\scriptscriptstyle 0T}}{\kappa_{\scriptscriptstyle 0C}}=T_0^3
\end{equation}
and
\begin{equation}
\frac{\varepsilon_2}{\varepsilon_1}=\frac{7+2\delta}{1+2\delta}\theta_0^3.\label{epsilon12}
\end{equation}

Applying the boundary condition of continuous heat flux (equivalent to continuous first derivative of temperature) at $\theta=\theta_0$, and substituting the above relations, we connect the first integrals of the two domains (Equations (\ref{ficoll}) and (\ref{fiturb})) at the interface point, by which the eigenvalue of $\xi$ is obtained and  explicitly expressed in a function of $\theta_0$ as
\begin{widetext}
\begin{equation}
\begin{aligned}
\xi&=\frac{6-(3-2\gamma+4\delta)\theta_0^{(3+2\gamma-4\delta)/2}-(3+2\gamma-4\delta)(\theta_b/\theta_0)^{(3-2\gamma+4\delta)/2}}{6-(7+2\alpha+4\delta)\theta_0^{-(1+2\alpha+4\delta)/2}+(1+2\alpha+4\delta)(\theta_b/\theta_0)^{(7+2\alpha+4\delta)/2}}\\
&\ \times\frac{(7+2\alpha+4\delta)(1+2\alpha+4\delta)}{(3-2\gamma+4\delta)(-3-2\gamma+4\delta)}\theta_0^{-(2+\gamma+\alpha)}.\label{xi}
\end{aligned}
\end{equation}
\end{widetext}

Once the value of $\xi$ is determined, the two first integrals can be directly integrated within their own domains, with the results ($\eta$ and $\zeta$, which are eventually converted back to $\theta$) expressed in implicit functions of $x$ as
\begin{equation}
\begin{aligned}
x&=\sqrt{\frac{\varepsilon_1}{2}}\int_{\eta_b}^{\eta}\left[\frac{\eta^{\mu_1+1}-\eta_b^{\mu_1+1}}{\mu_1+1}-\frac{\xi(\eta^{\nu_1+1}-\eta_b^{\nu_1+1})}{\nu_1+1}\right]^{-1/2}d\eta\\
&=\sqrt{\frac{\varepsilon_1}{2}}I_1(\eta(\theta))=\sqrt{\frac{\varepsilon_1}{2}}I_1(\theta)\label{xeta}
\end{aligned}
\end{equation}
and
\begin{equation}
\begin{aligned}
1-x&=\sqrt{\frac{\varepsilon_2}{2}}\int_{\zeta}^1\left[-\frac{1-\zeta^{\mu_2+1}}{\mu_2+1}+\frac{\xi(1-\zeta^{\nu_2+1})}{\nu_2+1}\right]^{-1/2}d\zeta\\
&=\sqrt{\frac{\varepsilon_2}{2}}I_2(\zeta(\theta))=\sqrt{\frac{\varepsilon_2}{2}}I_2(\theta),\label{xzeta}
\end{aligned}
\end{equation}
where we define the integrals
\begin{equation}
I_1(\theta)=\int_{\eta_b}^{\eta}\left[\frac{\eta^{\mu_1+1}-\eta_b^{\mu_1+1}}{\mu_1+1}-\frac{\xi(\eta^{\nu_1+1}-\eta_b^{\nu_1+1})}
{\nu_1+1}\right]^{-1/2}d\eta\label{I1theta}
\end{equation}
and
\begin{equation}
\begin{aligned}
I_2(\theta)=\int_{\zeta}^1\left[-\frac{1-\zeta^{\mu_2+1}}{\mu_2+1}+\frac{\xi(1-\zeta^{\nu_2+1})}{\nu_2+1}\right]^{-1/2}d\zeta,\label{I2theta}
\end{aligned}
\end{equation}
respectively.

Applying Equations (\ref{xeta}) and (\ref{xzeta}) at $x=\ell$ yields
\begin{equation}
\ell=\sqrt{\frac{\varepsilon_1}{2}}I_1(\theta_0)
\end{equation}
and
\begin{equation}
1-\ell=\sqrt{\frac{\varepsilon_2}{2}}I_2(\theta_0).
\end{equation}
Further applying the relation between $\varepsilon_1$ and $\varepsilon_2$ (Equation (\ref{epsilon12})), it is obtained that
\begin{equation}
\ell=\frac{I_1(\theta_0)}{I_1(\theta_0)+\left[(7+2\delta)/(1+2\delta)\right]^{1/2}\theta_0^{3/2}I_2(\theta_0)}=\frac{I_1(\theta_0)}{I(\theta_0)},\label{lratio}
\end{equation}
\begin{equation}
\varepsilon_1=\frac{2\ell^2}{I_1^2(\theta_0)}=\frac{2}{I^2(\theta_0)},\label{epsilon1}
\end{equation}
and
\begin{equation}
\varepsilon_2=\frac{2(7+2\delta)}{1+2\delta}\frac{\theta_0^3}{I^2(\theta_0)},
\end{equation}
where we make the identification
\begin{equation}
I(\theta_0)=I_1(\theta_0)+\left(\frac{7+2\delta}{1+2\delta}\right)^{1/2}\theta_0^{3/2} I_2(\theta_0)\label{Itheta0}
\end{equation}
for simplicity.

With the obtained eigenvalues of $\xi$, $\varepsilon_1$, and $\varepsilon_2$, the temperature profile of the loop is now completely determined, which is implicitly formulated in a piecewise form of
\begin{equation}
x=\left\{\begin{array}{cl}
\displaystyle \frac{I_1(\theta)}{I(\theta_0)};   & \theta_b\le\theta\le\theta_0\\
\displaystyle 1-\left(\frac{7+2\delta}{1+2\delta}\right)^{1/2}\frac{\theta_0^{3/2}}{I(\theta_0)}I_2(\theta);\ \   & \theta_0\le\theta\le1.\label{tprof}
\end{array}
\right.
\end{equation}
Furthermore, by translating the dimensionless eigenvalues to dimensional ones (see Equation (\ref{epsilonxi})), we obtain the loop-top temperature and heating rate as 
\begin{equation}
T_a=\left[\frac{(7+2\delta)\varepsilon_1}{8}\frac{\chi_0}{k_{\scriptscriptstyle\! B}^2\kappa_{\scriptscriptstyle 0C}}\right]^{2/(11+2\gamma)}\left(PL\right)^{4/(11+2\gamma)}\label{slta}
\end{equation}
and
\begin{equation}
H_a=\frac{\xi\chi_0}{4k_{\scriptscriptstyle\! B}^2}\frac{P^2}{T_a^{2+\gamma}}.\label{slha}
\end{equation}
These two equations will serve as a basis for the following deduction of loop scaling laws in various situations.

\section{Scaling Laws for Various Situations}\label{sec3}
Based on the generalized hybrid model, loop scaling laws of different forms are constructed for various situations. Here we set a zero loop-base temperature for analytical convenience. Since  $T_b\ll T_a$ holds for typical coronal loops, the specification of $\theta_b=0$ ($\eta_b=0$) is physically reasonable.

\subsection{The Collision-dominated Conduction Limit}\label{sec31}
In case of sufficiently strong collisions, the collision-dominated part will be lengthened to occupy the entire loop, which raises the value of  $\theta_0$ to unity. At this CDC limit, the hybrid model reduces to the single collisional model as previously explored in \citet{Martens10}. 

Since $\theta_0=1$, the originally complex expression of $\xi$ (Equation (\ref{xi})) becomes
\begin{equation}
\xi=\frac{7+2\alpha+4\delta}{3-2\gamma+4\delta}=\frac{\nu_1+1}{\mu_1+1}, \ \ \ [\mathrm{C}] 
\end{equation}
and the definite integral $I(\theta_0)$ (Equation (\ref{Itheta0})) consequently reduces to
\begin{equation}
\begin{aligned}
I(1)&=I_1(1)=\int_0^1\left(\frac{\eta^{\mu_1+1}-\eta^{\nu_1+1}}{\mu_1+1}\right)^{-1/2}d\eta\\
&=\frac{(\mu_1+1)^{1/2}}{\nu_1-\mu_1}B\left(\frac{1-\mu_1}{2(\nu_1-\mu_1)},\frac{1}{2}\right)\\
&=\frac{[(7+2\delta)(3-2\gamma+4\delta)]^{1/2}}{2(2+\gamma+\alpha)}\\
&\ \times B\left(\frac{11+2\gamma}{4(2+\gamma+\alpha)},\frac{1}{2}\right)\\
&=\left(\frac{7+2\delta}{4\phi}\right)^{1/2},\ \ \ [\mathrm{C}]\label{ic1}
\end{aligned}
\end{equation}
where $B$ is the complete beta function, and we introduce the auxiliary parameter
\begin{equation}
\phi=\frac{(2+\gamma+\alpha)^2}{3-2\gamma+4\delta}\left[B\left(\frac{11+2\gamma}{4(2+\gamma+\alpha)},\frac{1}{2}\right)\right]^{-2}\label{phi}
\end{equation}
for simplicity.

With the above value of $I(1)$, Equation (\ref{epsilon1}) becomes
\begin{equation}
\varepsilon_1=\frac{8\phi}{7+2\delta},\ \ \ [\mathrm{C}] 
\end{equation}
and Equation (\ref{tprof}) turns to
\begin{equation}
x=\beta_r\left(\theta^{2+\gamma+\alpha};\frac{11+2\gamma}{4(2+\gamma+\alpha)},\frac{1}{2}\right),\ \ \ [\mathrm{C}]
\end{equation}
which represents the temperature profile in a regularized incomplete beta function $\beta_r$.

Substituting the values of $\varepsilon_1$ and $\xi$ into Equations (\ref{slta}) and (\ref{slha}) gives
\begin{equation}
T_a=\left(\frac{\phi\chi_0}{k_{\scriptscriptstyle\! B}^2\kappa_{\scriptscriptstyle 0C}}\right)^{2/(11+2\gamma)}\left(PL\right)^{4/(11+2\gamma)}\ \ \ [\mathrm{C}]\label{sltac} 
\end{equation} 
and
\begin{equation}
\begin{aligned}
H_a&=\frac{7+2\alpha+4\delta}{4(3-2\gamma+4\delta)\phi}\left(\frac{\phi\chi_0}{k_{\scriptscriptstyle\! B}^2\kappa_{\scriptscriptstyle 0C}}\right)^{7/(11+2\gamma)}\\
&\ \times\kappa_{\scriptscriptstyle 0C} P^{14/(11+2\gamma)}L^{-4(2+\gamma)/(11+2\gamma)},\ \ \ [\mathrm{C}]\label{slhac} 
\end{aligned}
\end{equation}
which are the generalized loop scaling laws for the CDC limit. Note that the above formalism is largely the same as that in \citet{Martens10}. Here we mark the relevant equations with a notation of ``C" that stands for collisional situation.

Setting $\chi_0=1.55\times10^{-19}$ erg cm$^3$ s$^{-1}$ K$^{1/2}$ and $\gamma=1/2$ in the single power-law radiation function \citep[e.g.,][]{Cox69,Rosner78}, the CDC scaling laws for the specific  case of $\alpha=0$ (uniform heating)  and $\delta=0$ (constant loop cross section) are cast to 
\begin{equation}
T_a\approx1.3\times10^3(PL)^{1/3}\ \ \mathrm{K}
\end{equation}
and
\begin{equation}
H_a(=E_{\scriptscriptstyle\! H})\approx1.2\times10^5P^{7/6}L^{-5/6} \ \ \mathrm{erg}\ \mathrm{cm}^{-3}\ \mathrm{s}^{-1},
\end{equation}
which correspond to the well-known RTV scaling laws \citep[with slight differences in proportional coefficients from the original equations in][]{Rosner78}.

It is worth pointing out that a valid definition of the beta function in Equation (\ref{ic1}) requires that $(11+2\gamma)/4(2+\gamma+\alpha)>0$, namely, $\alpha>-(2+\gamma)$. It means that an equilibrium loop solution exists only when the background heating is not too concentrated near the loop base.  Otherwise, the equilibrium will be broken and the loop will enter into a state of thermal nonequilibrium \citep[TNE,][]{Klimchuk10,Froment18,Klimchuk19}.
 
At the CDC limit, we combine Equations (\ref{eqnlambc}) and (\ref{sltac}) to evaluate the (maximum) collisional mean free path at the loop apex, which gives a critical value of 
\begin{equation}
\lambda_{Tc}=\left(\frac{\phi\chi_0}{k_{\scriptscriptstyle\! B}^2\kappa_{\scriptscriptstyle 0C}}\right)^{6/(11+2\gamma)}c_{\scriptscriptstyle\! R}\left(PL\right)^{(1-2\gamma)/(11+2\gamma)}L\label{lambtc}
\end{equation}
for the turbulent mean free path. Obviously the CDC limit holds for $\lambda_T\ge\lambda_{Tc}$.

\subsection{The Turbulence-dominated Conduction Limit}\label{sec32}
 At the limit of $\lambda_T\to0$ ($\lambda_T\ll\lambda_{Tc}$), on the contrary, the (dimensionless) interface temperature approaches zero, and most part of the loop is dominated by turbulent scattering.\footnote[3]{The value of $T_0$ ($\theta_0$) cannot decrease to zero, otherwise a singularity will occur at the footpoint of the loop. This means that a loop with a zero loop-base temperature must host a collision-dominated part, although its extent could be extremely small. The constraint is relaxed in case of a nonzero loop-base temperature, where the loop could be wholly collision-dominated provided that the turbulent mean free path becomes shorter than the collisional mean free path evaluated at the loop base \citep[see,][]{Bradshaw19}.}  Combining Equations (\ref{lambt}) and (\ref{lambtc}) with the aid of Equations (\ref{epsilon1}) and (\ref{slta}), we can relate the normalized turbulent mean free path $\lambda_T/\lambda_{Tc}$ (when $\lambda_T/\lambda_{Tc}\le1$) to $\theta_0$ by
\begin{equation}
\frac{\lambda_T}{\lambda_{Tc}}=\left[\frac{7+2\delta}{4\phi I^2(\theta_0)}\right]^{6/(11+2\gamma)}\theta_0^3,\label{theta0lambt}
\end{equation}
based on which loop scaling laws for the TDC limit could be built in functions of $\lambda_T$.

At the TDC limit, the value of the definite integral $I(\theta_0)$ should be predominately determined by the turbulent part, and is therefore approximated as 
\begin{equation}
I(\theta_0)\approx\left(\frac{7+2\delta}{1+2\delta}\right)^{1/2}\theta_0^{3/2} I_2(\theta_0).\label{Itheta0t}
\end{equation}
Moreover, unlike the collisional integral $I_1(\theta_0)$ (see Equation (\ref{I1theta})) where both $\mu_1+1=(3-2\gamma+4\delta)/(7+2\delta)$ and $\nu_1+1 =(7+2\alpha+4\delta)/(7+2\delta)$ are greater than zero, in the turbulent integral $I_2(\theta_0)$ (see Equation (\ref{I2theta})) , the indices $\mu_2+1=(-3-2\gamma+4\delta)/(1+2\delta)$ and $\nu_2+1=(1+2\alpha+4\delta)/(1+2\delta)$ can be both greater than zero, both less than zero, or of opposite signs. Therefore, with different specifications of the $\gamma$, $\alpha$, and $\delta$ values, the expression of $\xi$ may have different functional forms as $\theta_0$ asymptotically approaches zero, which will consequently lead to to different forms of scaling laws.

The first situation is $\mu_2+1<\nu_2+1<0$, namely, $\alpha\in\left(-(2+\gamma),\, -(1+4\delta)/2\right)$, which corresponds to a nonuniform heating moderately concentrated near the loop footpoint. Under this condition, 
\begin{equation}
\xi\approx\frac{(7+2\alpha+4\delta)(1+2\alpha+4\delta)}{(3-2\gamma+4\delta)(-3-2\gamma+4\delta)}\theta_0^{-(2+\gamma+\alpha)}\ \ \ [\mathrm{T1}]\label{xit1} 
\end{equation}
for small values of $\theta_0$ (from Equation (\ref{xi})). Since $\xi\propto\zeta_0^{\mu_2-\nu_2}\to\infty$ as $\theta_0\to0$, the integrand in $I_2(\theta_0)$ is dominated by the term involving $\xi$, and meanwhile the lower bound of the integral can be tactically substituted with zero. Therefore, Equation (\ref{Itheta0t}) becomes
\begin{equation}
\begin{aligned}
I(\theta_0)&\approx\left(\frac{7+2\delta}{1+2\delta}\right)^{1/2}{\theta_0^{3/2}}\int_{\zeta_0}^{1}\left[\frac{\xi(1-\zeta^{\nu_2+1})}{\nu_2+1}\right]^{-1/2}d\zeta\\
&\approx\left(\frac{7+2\delta}{1+2\delta}\right)^{1/2}\frac{\theta_0^{3/2}}{\xi^{1/2}}\int_0^1\left(\frac{1-\zeta^{\nu_2+1}}{\nu_2+1}\right)^{-1/2}d\zeta\\
&=\left[-\frac{7+2\delta}{(1+2\delta)(\nu_2+1)}\right]^{1/2}B\left(\frac{\nu_2-1}{2(\nu_2+1)},\frac{1}{2}\right)\frac{\theta_0^{3/2}}{\xi^{1/2}}\\
&=\left(\frac{7+2\delta}{4\phi_1}\right)^{1/2}\theta_0^{(5+\gamma+\alpha)/2},\ \ \ [\mathrm{T1}]\label{Itheta0t1} 
\end{aligned}
\end{equation}
and Equation (\ref{epsilon1}) becomes
\begin{equation}
\varepsilon_1\approx\frac{8\phi_1}{7+2\delta}\theta_0^{-(5+\gamma+\alpha)},\ \ \ [\mathrm{T1}]\label{epsilont1}
\end{equation}
where we define the auxiliary parameter
\begin{equation}
\begin{aligned}
\phi_1&=\frac{(7+2\alpha+4\delta)(1+2\alpha+4\delta)^2}{4(3-2\gamma+4\delta)(3+2\gamma-4\delta)}\\
&\ \times\left[B\left(\frac{-1+2\alpha}{2(1+2\alpha+4\delta)},\frac{1}{2}\right)\right]^{-2}
\end{aligned}
\end{equation}
for simplicity.

Relating Equation (\ref{Itheta0t1}) to Equation (\ref{theta0lambt}) yields 
\begin{equation}
\theta_0\approx\left(\frac{\phi_1}{\phi}\right)^{-2/(1-2\alpha)}\left(\frac{\lambda_T}{\lambda_{Tc}}\right)^{(11+2\gamma)/3(1-2\alpha)},\ \ \ [\mathrm{T1}] 
\end{equation}
which reveals a power-law dependence of $\theta_0$ on $\lambda_T$. Using this relation to substitute $\theta_0$ in Equations (\ref{xit1}) and (\ref{epsilont1}), and inserting the results into Equations (\ref{slta}) and (\ref{slha}), the TDC scaling laws for the case of moderately stratified footpoint heating are formulated as
\begin{equation}
\begin{aligned}
T_a& \approx\left(\frac{\phi_1}{\phi}\right)^{2/(1-2\alpha)}\left(\frac{\phi\chi_0}{k_{\scriptscriptstyle\! B}^2\kappa_{\scriptscriptstyle 0C}}\right)^{2/(11+2\gamma)}\left(PL\right)^{4/(11+2\gamma)}\\
&\ \times\left(\frac{\lambda_T}{\lambda_{Tc}}\right)^{-2(5+\gamma+\alpha)/3(1-2\alpha)}\ \ \ [\mathrm{T1}]\label{sltat1} 
\end{aligned}
\end{equation}
and
\begin{equation}
\begin{aligned}
H_a&\approx\frac{(7+2\alpha+4\delta)(1+2\alpha+4\delta)}{4(3-2\gamma+4\delta)(-3-2\gamma+4\delta)\phi}\left(\frac{\phi_1}{\phi}\right)^{2\alpha/(1-2\alpha)}\\
&\ \times\left(\frac{\phi\chi_0}{k_{\scriptscriptstyle\! B}^2\kappa_{\scriptscriptstyle 0C}}\right)^{7/(11+2\gamma)} P^{14/(11+2\gamma)}L^{-4(2+\gamma)/(11+2\gamma)}\\
&\ \times \kappa_{\scriptscriptstyle 0C}\left(\frac{\lambda_T}{\lambda_{Tc}}\right)^{-(2+\gamma+7\alpha)/3(1-2\alpha)}.\ \ \ [\mathrm{T1}] \label{slhat1}
\end{aligned}
\end{equation}

The second situation is $\mu_2+1<0<\nu_2+1$, which is equivalent to $\delta\in\left(-(1+2\alpha)/4,\, (3+2\gamma)/4\right)\cap[0,\, +\infty)$. It physically implies a moderate expansion of the loop cross section with height or uniform cross-sectional area throughout the loop (only if $\alpha>-1/2$). For this situation,
\begin{equation}
\xi\approx\frac{6(1+2\alpha+4\delta)}{(3-2\gamma+4\delta)(3+2\gamma-4\delta)}\theta_0^{-(3+2\gamma-4\delta)/2},\ \ \ [\mathrm{T2}]\label{xit2}
\end{equation}
which also approaches infinity with the decrease of $\theta_0$ but in another power-law form proportional to $\zeta_0^{\mu_2+1}$. With the similar approximations as made in the first turbulent situation, the definite integral $I(\theta_0)$ is accordingly evaluated as
\begin{equation}
\begin{aligned}
I(\theta_0)&\approx\left(\frac{7+2\delta}{1+2\delta}\right)^{1/2}\frac{\theta_0^{3/2}}{\xi^{1/2}}\int_0^1\left(\frac{1-\zeta^{\nu_2+1}}{\nu_2+1}\right)^{-1/2}d\zeta\\
&=\left[\frac{7+2\delta}{(1+2\delta)(\nu_2+1)}\right]^{1/2}B\left(\frac{1}{\nu_2+1},\frac{1}{2}\right)\frac{\theta_0^{3/2}}{\xi^{1/2}}\\
&=\left(\frac{7+2\delta}{4\phi_2}\right)^{1/2}\theta_0^{(9+2\gamma-4\delta)/4},\ \ \ [\mathrm{T2}] \label{Itheta0t2} 
\end{aligned}
\end{equation}
where we make the identification 
\begin{equation}
\begin{aligned}
\phi_2&=\frac{3(1+2\alpha+4\delta)^2}{2(3-2\gamma+4\delta)(3+2\gamma-4\delta)}\\
&\ \times\left[B\left(\frac{1+2\delta}{1+2\alpha+4\delta},\frac{1}{2}\right)\right]^{-2}
\end{aligned}
\end{equation}
for simplicity. Combining Equations (\ref{theta0lambt}) and (\ref{Itheta0t2}), the dependence between $\lambda_T$ and $\theta_0$ then changes to
\begin{equation}
\begin{aligned}
\theta_0&\approx\left(\frac{\phi_2}{\phi}\right)^{-1/(1+2\delta)}\left(\frac{\lambda_T}{\lambda_{Tc}}\right)^{(11+2\gamma)/6(1+2\delta)}.\ \ \ [\mathrm{T2}] 
\end{aligned}
\end{equation}
Consequently, the TDC scaling laws for the case of moderate loop cross-sectional expansion turn out to be
\begin{equation}
\begin{aligned}
T_a&\approx\left(\frac{\phi_2}{\phi}\right)^{1/(1+2\delta)}\left(\frac{\phi\chi_0}{k_{\scriptscriptstyle\! B}^2\kappa_{\scriptscriptstyle 0C}}\right)^{2/(11+2\gamma)}\left(PL\right)^{4/(11+2\gamma)}\\
&\ \times\left(\frac{\lambda_T}{\lambda_{Tc}}\right)^{-(9+2\gamma-4\delta)/6(1+2\delta)}\ \ \ [\mathrm{T2}]\label{sltat2}  
\end{aligned}
\end{equation}
and
\begin{equation}
\begin{aligned}
H_a&\approx\frac{3(1+2\alpha+4\delta)}{2(3-2\gamma+4\delta)(3+2\gamma-4\delta)\phi}\left(\frac{\phi_2}{\phi}\right)^{-(1+4\delta)/2(1+2\delta)}\\
&\ \times\left(\frac{\phi\chi_0}{k_{\scriptscriptstyle\! B}^2\kappa_{\scriptscriptstyle 0C}}\right)^{7/(11+2\gamma)}P^{14/(11+2\gamma)}L^{-4(2+\gamma)/(11+2\gamma)}\\
&\ \times \kappa_{\scriptscriptstyle 0C}\left(\frac{\lambda_T}{\lambda_{Tc}}\right)^{(3-2\gamma+28\delta)/12(1+2\delta)}.\ \ \ [\mathrm{T2}] \label{slhat2} 
\end{aligned}
\end{equation}

The third situation is $0<\mu_2+1<\nu_2+1$, that is, $\delta\in\left((3+2\gamma)/4,\, +\infty\right)$. In comparison with the second turbulent situation, the current one represents a relatively stronger expansion of the loop cross section. Under this condition,
\begin{equation}
\xi\approx\frac{1+2\alpha+4\delta}{-3-2\gamma+4\delta}=\frac{\nu_2+1}{\mu_2+1},\ \ \ [\mathrm{T3}] 
\end{equation}
which asymptotically approaches a constant value for small values of $\theta_0$. With the approximately constant $\xi$ value, all the terms in the integrand of $I_2(\theta_0)$ are comparable, and hence the definite integral $I(\theta_0)$ is evaluated in a similar way to that for the collisional case, which yields
\begin{equation}
\begin{aligned}
I(\theta_0)&\approx\left(\frac{7+2\delta}{1+2\delta}\right)^{1/2}\theta_0^{3/2}\int_0^1\left(\frac{\zeta^{\mu_2+1}-\zeta^{\nu_2+1}}{\mu_2+1}\right)^{-1/2}d\zeta\\
&=\frac{\left[(7+2\delta)(\mu_2+1)\right]^{1/2}}{(1+2\delta)^{1/2}(\nu_2-\mu_2)}B\left(\frac{1-\mu_2}{2(\nu_2-\mu_2)},\frac{1}{2}\right)\theta_0^{3/2}\\
&=\left(\frac{7+2\delta}{4\phi_3}\right)^{1/2}\theta_0^{3/2},\ \ \ [\mathrm{T3}] \label{Itheta0t3} 
\end{aligned}
\end{equation}
where we introduce the auxiliary parameter
\begin{equation}
\phi_3=\frac{(2+\gamma+\alpha)^2}{-3-2\gamma+4\delta}\left[B\left(\frac{5+2\gamma}{4(2+\gamma+\alpha)},\frac{1}{2}\right)\right]^{-2}
\end{equation}
for simplicity. Combining Equations (\ref{theta0lambt}) and (\ref{Itheta0t3}), it is obtained that
\begin{equation}
\theta_0\approx\left(\frac{\phi_3}{\phi}\right)^{-2/(5+2\gamma)}\left(\frac{\lambda_T}{\lambda_{Tc}}\right)^{(11+2\gamma)/3(5+2\gamma)}.\ \ \ [\mathrm{T3}]
\end{equation}
With this relation, Equations (\ref{slta}) and (\ref{slha}) finally reduce to
\begin{equation}
\begin{aligned}
T_a&\approx\left(\frac{\phi\chi_0}{k_{\scriptscriptstyle\! B}^2\kappa_{\scriptscriptstyle 0C}}\right)^{2/(11+2\gamma)}\left(PL\right)^{4/(11+2\gamma)}\\
&\ \times\left(\frac{\phi}{\phi_3}\frac{\lambda_T}{\lambda_{Tc}}\right)^{-2/(5+2\gamma)}\ \ \ [\mathrm{T3}]\label{sltat3} 
\end{aligned}
\end{equation}
and
\begin{equation}
\begin{aligned}
H_a&\approx\frac{1+2\alpha+4\delta}{4(-3-2\gamma+4\delta)\phi}\left(\frac{\phi\chi_0}{k_{\scriptscriptstyle\! B}^2\kappa_{\scriptscriptstyle 0C}}\right)^{7/(11+2\gamma)}\\
&\ \times\kappa_{\scriptscriptstyle 0C}P^{14/(11+2\gamma)}L^{-4(2+\gamma)/(11+2\gamma)}\\
&\ \times\left(\frac{\phi}{\phi_3}\frac{\lambda_T}{\lambda_{Tc}}\right)^{2(2+\gamma)/(5+2\gamma)},\ \ \ [\mathrm{T3}]\label{slhat3} 
\end{aligned}
\end{equation}
which give the TDC scaling laws for the case of strong loop cross-sectional expansion.

To discriminate the three different turbulent situations, we assign the relevant equations with notations of ``T1," ``T2," and ``T3," respectively. According to our classification, the case of $\gamma=1/2$, $\alpha=0$, and $\delta=0$ at the TDC limit belongs to situation T2. For this specific case, our Equations (\ref{sltat2}) and (\ref{slhat2}) naturally reduce to the scaling laws derived in \citet[][their Equations (46) and (47)]{Emslie22} after some algebra. It is worth pointing out that  their scaling laws involve both $\lambda_T$ and $T_0$, which are mutually dependent (see Equation (\ref{lambt})). With the relation between the two quantities taken into account, our formalism should be more physically straightforward.

Finally, with the approximate expressions of $\xi$ put into $I_1(\theta_0)$, it  is derived that $I_1(\theta_0)\propto\theta_0^{(11+2\gamma)/4}$ for all three turbulent situations. This results in ratios of $I_1(\theta_0)/I(\theta_0)$ (normalized length of the collision-dominated part according to Equation (\ref{lratio})) in proportion to $\theta_0^{(1-2\alpha)/4}$ (T1), $\theta_0^{(1+2\delta)/2}$ (T2), and $\theta_0^{(5+2\gamma)/4}$ (T3), respectively,  all of which approach zero at the TDC limit. Therefore, the validity of neglecting $I_1(\theta_0)$ in Equation (\ref{Itheta0t}) is a posterior verified.

\section{Effects of Nonuniform Heating and Cross Section on Turbulent Loops}\label{sec4}
\subsection{Turbulent Solutions for Uniform Heating and Cross Section}
\begin{figure*}
\epsscale{0.85}
\plotone{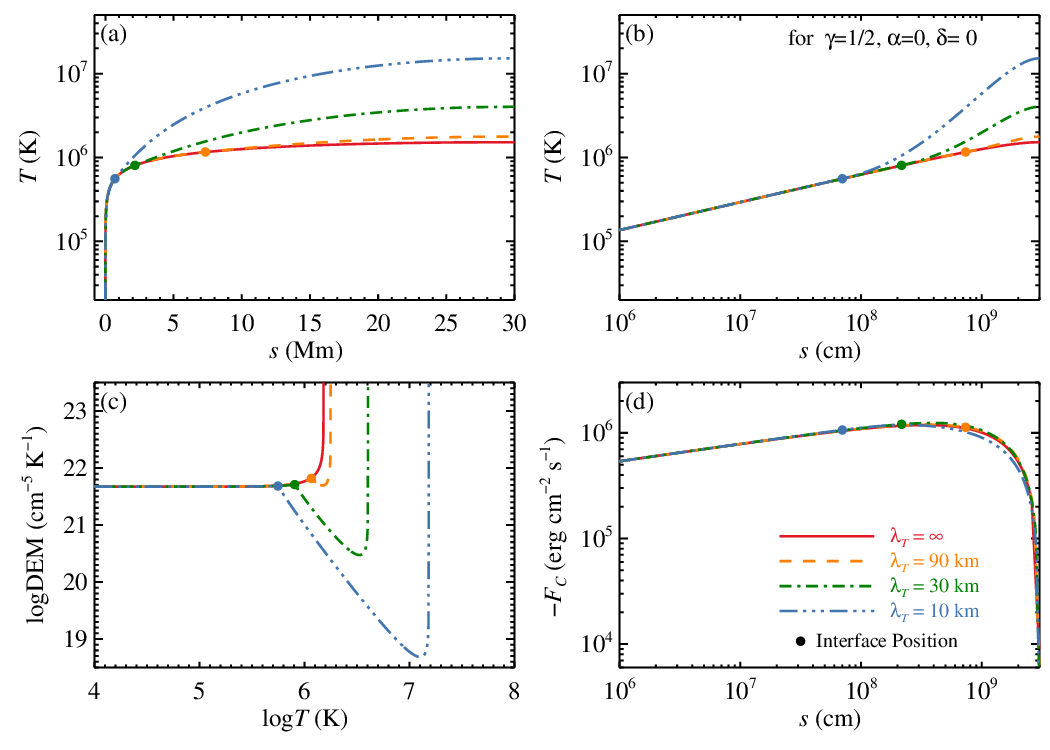}
\caption{Hybrid model solutions of an isobaric coronal loop computed for different turbulent mean free paths (discriminated with different colors and line styles), with the panels showing profiles of loop temperature (a and b), DEM (c), and heat flux (d), respectively. The loop has a half-length of 30 Mm and constant pressure of 0.55 dyne cm$^{-2}$. All the cases are powered by a spatially uniform heating ($\alpha=0$) and have a constant cross section ($\delta=0$). The solid circle outlines the interface between the collision-dominated part and the turbulence-dominated part for each case.
\label{fig01}}
\end{figure*}

\begin{figure*}
\epsscale{0.7}
\plotone{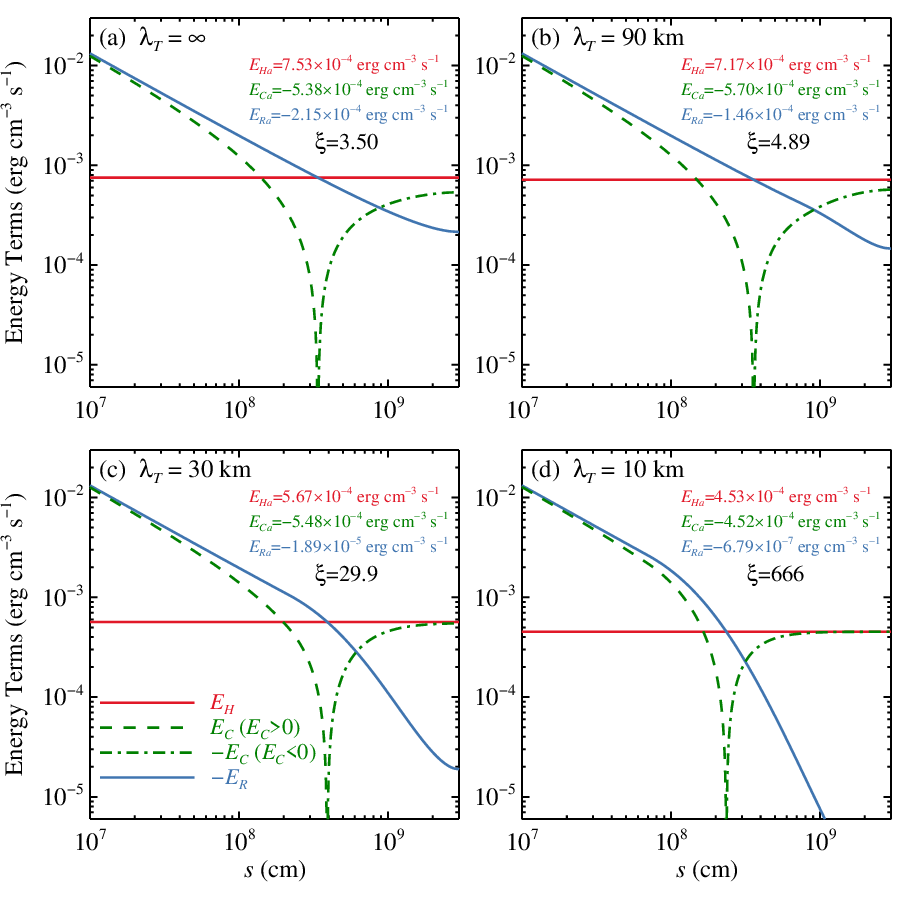}
\caption{Magnitudes of energy terms  for the same loop modeled in Figure \ref{fig01}, with different panels demonstrating the cases of different turbulent mean free paths. In each panel, the red, green, and blue lines represent terms of heating ($E_{\scriptscriptstyle\! H}$), conduction~($E_{\scriptscriptstyle\! C}$), and radiation ($E_{\scriptscriptstyle\! R}$), respectively. In plotting the thermal conduction, we adopt different line styles to trace its transition from an energy loss term (dash-dotted) to a gain term (dashed).
\label{fig02}}
\end{figure*}

Before exploring the effect of  nonuniform heating and loop cross section on turbulent loop solutions, it is necessary to first address how turbulent scattering affects a uniformly heated loop with a constant cross section. Figure \ref{fig01} plots such hybrid model solutions of an isobaric coronal loop computed for different turbulent mean free paths. The loop has a half-length of 30~Mm and  pressure of 0.55 dyne~cm$^{-2}$, typical of an AR loop. As for the single power-law formed radiation function, we set the index $\gamma$ at a commonly adopted value of 1/2.  These parameters result in a loop-top temperature of 1.52~MK for the wholly collisional case (e.g., $\lambda_T=\infty$), as well as a critical turbulent mean free path of $\lambda_{Tc}=202$~km.

Compared with the wholly collisional case, the inclusion of turbulent scattering (when $\lambda_T<\lambda_{Tc}$) increases the temperature in the turbulence-dominated part, while in regions below the interface point, the temperature profile is almost unchanged (Figures \ref{fig01}a and \ref{fig01}b). This temperature enhancement leads to both a decrease of the density (to keep the constant loop pressure) and a sharpening of the temperature gradient in the corona. As a result, the DEM profile, defined as $\mathrm{DEM}(T)=n_e^2(dT/ds)^{-1}$, reveals a depletion in the corona relative to that in the TR, especially for strong turbulence (Figure \ref{fig01}c). Note that based on the hybrid model solutions, the DEM in the collision-dominated part is readily formulated as
\begin{equation}
\begin{aligned}
\mathrm{DEM}(T)&=\frac{(3-2\gamma+4\delta)^{1/2}}{4k_{\scriptscriptstyle\! B}}\left(\frac{\kappa_{\scriptscriptstyle 0C}}{\chi_0}\right)^{1/2}PT^{-(1-2\gamma)/4}\\
&\ \times\left[1-\frac{3-2\gamma+4\delta}{7+2\alpha+4\delta}\xi \left(\frac{T}{T_a}\right)^{2+\gamma+\alpha}\right]^{-1/2}.
\end{aligned}
\end{equation}
With a $\gamma$ value of 1/2 and the same specification of $\delta$,  all DEM profiles will keep nearly the same flat level at $\left((1+2\delta)\kappa_{\scriptscriptstyle 0C}/8\chi_0k_{\scriptscriptstyle\! B}^2\right)^{1/2}P$ for low enough temperatures, as seen in Figure \ref{fig01}c as well as in \citet[][their Figure 4]{Emslie22}.

The effect of turbulence can be understood as follows. In case of TDC, the thermal conduction coefficient has a much weaker (or even inverse) dependence on temperature ($\kappa\propto T^{-1/2}$, see Equation(\ref{kappatdc})) than that for CDC ($\kappa\propto T^{5/2}$, see Equation(\ref{kappacdc})), leading to a suppression of the conduction coefficient in this regime. When turbulent scattering sets in, the temperature gradient in the turbulence-dominated part must be accordingly sharpened to maintain a relatively unchanged heat flux (compared to the wholly collisional situation). As shown in Figure \ref{fig01}d, the profiles of heat flux calculated for the different turbulent mean free paths do closely resemble each other over the entire loop, in sharp contrast to the wide divergence of the coronal temperature profiles.

Figure \ref{fig02} displays the magnitudes of energy terms for the same loop modeled in Figure \ref{fig01}. In the corona, the balance of energy is primarily between volumetric heating and conductive losses. As expected, turbulent scattering significantly reduces the radiative losses around the loop apex, but only marginally affects the thermal conduction. This leads to a slight depression of the loop heating rate, and meanwhile the ratio of loop-top heating to radiation, i.e., the dimensionless parameter $\xi$ (from Equation (\ref{epsilonxi})), increases significantly with the decrease of $\lambda_T$. In passing we note that to the lower part of the loop, the energy balance becomes predominately between conductive gain and radiative losses. In this sense, even a notable variation of the loop heating could barely influence the shape of temperature profile in this region (see Figures \ref{fig01}a and \ref{fig01}b). 

Although the inclusion of strong enough turbulent scattering can cause a relative excess of emitting materials in the TR to those in the corona, more compatible with observations than the wholly collisional model, it also significantly elevates the coronal temperature in case of uniform heating and cross section. For the case of $\lambda_T=10$~km shown in Figure \ref{fig01}, the loop apex has reached an unrealistically high temperature over 15~MK, which seems to impose a big challenge on the viability of such strength of turbulence working in nonflaring AR loops. However, as we will see, a nonuniform heating and/or cross section may substantially alter the dependence of loop temperature on turbulence strength.

\subsection{Effects of Nonuniform Heating and Cross Section}
\begin{figure*}
\epsscale{0.95}
\plotone{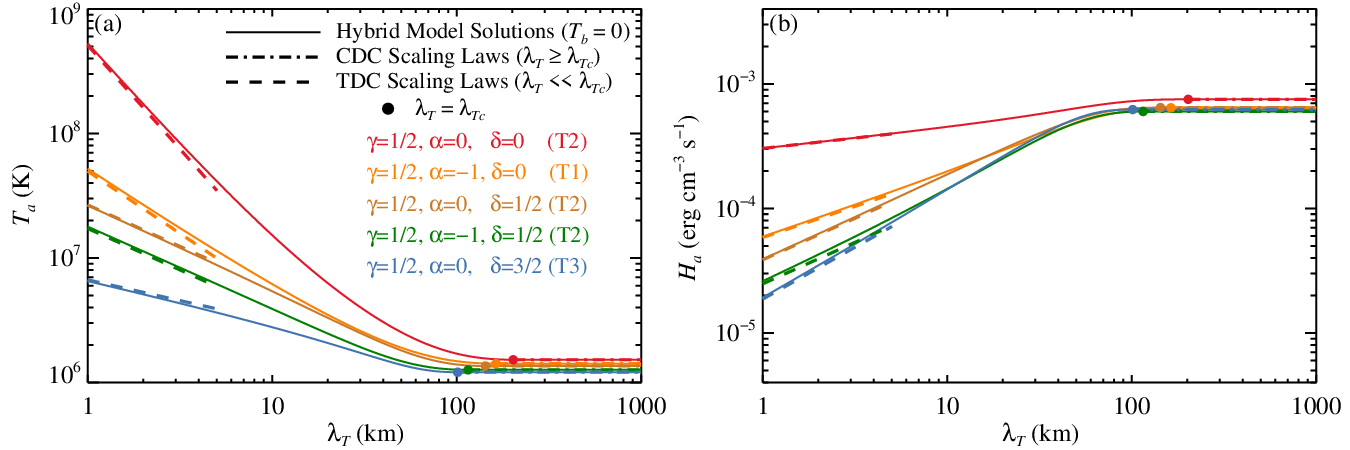}
\caption{Variations of loop-top temperature (left) and heating rate (right) vs.\ turbulent mean free path for an isobaric coronal loop with a half-length of 30 Mm and pressure of 0.55 dyne cm$^{-2}$.  The solid, dash-dotted, and dashed lines denote the loop quantities evaluated with the hybrid model solutions, CDC scaling laws, and TDC scaling laws, respectively. Different colors represent cases of different heating and/or areal parameters, which involve various turbulent situations (indicated in the legend). The solid circle depicts point of the critical turbulent mean free path for each case.\label{fig03}}
\end{figure*}

Figure \ref{fig03} plots the variations of loop-top temperature and heating rate versus turbulent mean free path for an isobaric coronal loop with the same half-length and pressure as those adopted in Figure \ref{fig01}. Here we fix the $\gamma$ value at 1/2 as before, and vary the values of $\alpha$ and $\delta$ to cover all three turbulent situations. The loop properties are evaluated with the hybrid model solutions (solid lines),  CDC scaling laws (dash-dotted lines), and  TDC scaling laws (dashed lines), respectively. 

For $\lambda_T\ge\lambda_{Tc}$, the hybrid model solutions are equivalent to the CDC scaling laws (Equations (\ref{sltac}) and (\ref{slhac})), where the loop quantities are analytically determined irrespective of $\lambda_T$.  At the CDC limit, the loop quantities are not very sensitive to the specification of  $\alpha$ and/or $\delta$, at least for the cases plotted in the figure. Between the different cases, the values of $T_a$ and $H_a$ lie in narrow ranges of 1.21--1.52~MK and 6.01--7.53$\times10^{-4}$~erg~cm$^{-3}$~s$^{-1}$, respectively. Even the resultant critical mean free path ($\propto T_a^3$) just varies within a factor of two.

For $\lambda_T<\lambda_{Tc}$, in contrast, the evolutions of the loop quantities diverge significantly between the different cases. At small values of $\lambda_T$ (e.g., $\lambda_T<0.05\lambda_{Tc}$), this case-to-case divergence is well characterized by the TDC scaling laws (Equations (\ref{sltat1}--\ref{slhat1}), (\ref{sltat2}--\ref{slhat2}), and (\ref{sltat3}--\ref{slhat3})), which give a robust estimate of the hybrid model solutions but use a more simple and straightforward approach.

\begin{deluxetable*}{ccchc}
\tablecaption{Conditions and Power-law Indices for the Various TDC Scaling Laws}\label{tb1}
\tablehead{
\colhead{Mathematical} & \multicolumn{2}{c}{Power-law Index on $\lambda_T$} & \colhead{} & \colhead{Category} \\
\cline{2-3} 
\colhead{Condition} & \colhead{$T_a$} &  \colhead{$H_a$}  & \colhead{} & \colhead{Notation}}
\startdata
$\alpha\in\left(-(2+\gamma),\, -(1+4\delta)/2\right)$ & $-2(5+\gamma+\alpha)/3(1-2\alpha)$ & $-(2+\gamma+7\alpha)/3(1-2\alpha)$ & & \\ 
$\gamma=1/2,\alpha=-1,\delta=0$ & $-1$ & $1/2$ & &\\
$\gamma=1/2,\alpha=-3/2,\delta=0$ & $-2/3$ & $2/3$ & &\raisebox{2.3ex}[0pt]{T1} \\
$\gamma=1/2,\alpha=-3/2,\delta=1/4$ & $-2/3$ & $2/3$ & &\\
\hline
$\delta\in\left(-(1+2\alpha)/4,\, (3+2\gamma)/4\right)\cap[0,\, +\infty)$ & $-(9+2\gamma-4\delta)/6(1+2\delta)$ & $(3-2\gamma+28\delta)/12(1+2\delta)$  & & \\
$\gamma=1/2,\alpha=0,\delta=0$ & $-5/3$ & $1/6$ & & \\
$\gamma=1/2,\alpha=0,\delta=1/2$ & $-2/3$ & $2/3$ & &\raisebox{2.3ex}[0pt]{T2}\\
$\gamma=1/2,\alpha=-1,\delta=1/2$ & $-2/3$ & $2/3$ & &\\
\hline
$\delta\in\left((3+2\gamma)/4,\, +\infty\right)$ & $-2/(5+2\gamma)$ & $2(2+\gamma)/(5+2\gamma)$  & & \\
$\gamma=1/2,\alpha=0,\delta=3/2$ & $-1/3$ & $5/6$ & &\\
$\gamma=1/2,\alpha=-1,\delta=3/2$ & $-1/3$ & $5/6$ & &\raisebox{2.3ex}[0pt]{T3}\\
$\gamma=1/2,\alpha=0,\delta=2$ & $-1/3$ & $5/6$ & &
\enddata
\end{deluxetable*} 

Compared with the conventional CDC scaling laws, the TDC scaling laws are additionally power-law functions of $\lambda_T$, with the functional forms varying from situation to situation. In Table \ref{tb1} we categorize and list the power-law indices of the various TDC scaling laws, including their general forms as well as particular values computed with specified values of $\gamma$, $\alpha$, and $\delta$. For situation T1 (moderately stratified footpoint heating), once the radiative loss function is fixed, the power-law indices solely depend on $\alpha$, while the value of $\delta$  only influences the proportionality coefficients. For situation T2 (moderate loop cross-sectional expansion), the indices are determined by $\delta$  instead, with $\alpha$ just playing a role in the proportionality coefficients. It is found that both a decrease of sufficiently negative $\alpha$ (for situation T1) and an increase of $\delta$ (for situation T2) can remarkably weaken (strengthen) the negative (positive) power-law dependence of $T_a$ ($H_a$) on $\lambda_T$. It means that for a given turbulence strength, the involvement of either a footpoint  heating or a cross-sectional expansion or both may effectively lower the coronal temperature of a loop, as revealed in Figure~\ref{fig03}a. Here we emphasize on the behavior of $T_a$ rather than $H_a$, for the reason that loop temperature is a measurable quantity whereas background heating is not. The effect of loop cross-sectional expansion seems to saturate upon a certain degree. For situation T3 when $\delta$ is sufficiently large, neither $\alpha$ nor $\delta$  affects the power-law indices any more, and the dependence between $T_a$ and $\lambda_T$ becomes the weakest.

\begin{figure*}
\epsscale{0.85}
\plotone{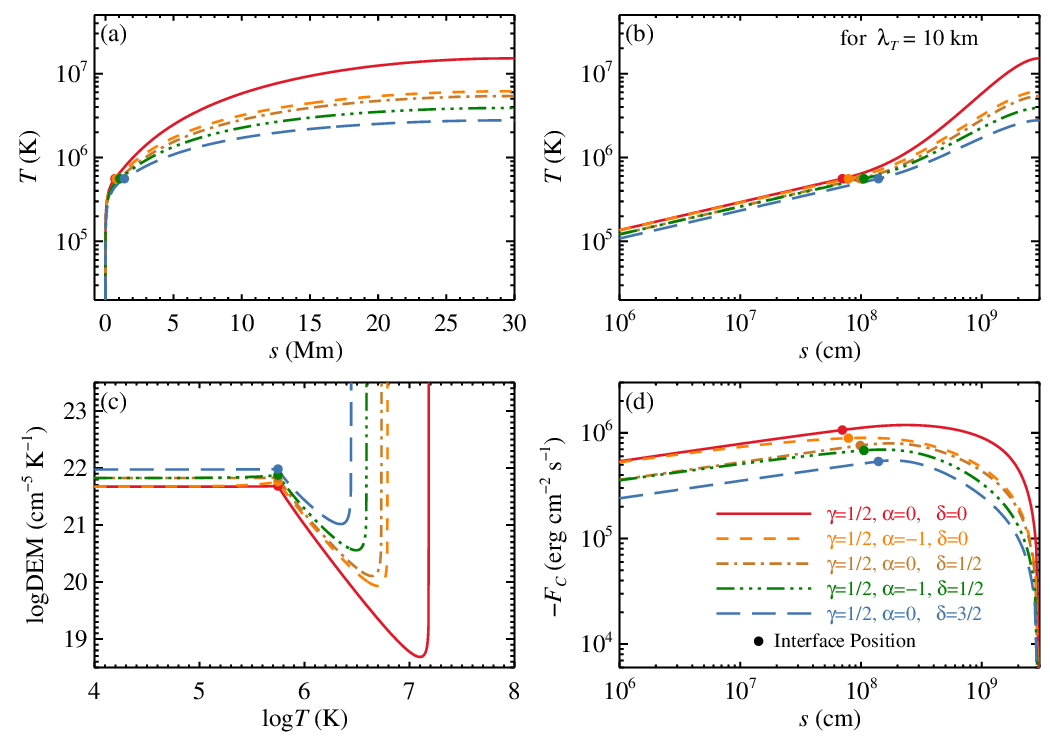}
\caption{The same as Figure \ref{fig01}, but for various heating ($\alpha$) and/or areal ($\delta$) parameters together with a fixed turbulent mean free path of 10~km.\label{fig04}}
\end{figure*}

Full loop solutions help better understand the effects of nonuniform heating and cross section. Figure \ref{fig04} plots hybrid model solutions of the loop computed for the various heating and/or areal parameters together with a fixed $\lambda_T$ value of 10~km. Compared with the case of uniform heating and cross section, it is seen that both a footpoint-concentrated heating and a cross-sectional expansion notably reduce the heat flux in the corona (Figure \ref{fig04}d). The region of reduced heat flux largely overlaps with the turbulence-dominated part, where turbulent scattering introduces a much weaker (or even inverse) dependence of the conduction coefficient on temperature. Therefore, the reduction of the heat flux in this region must be realized predominately by flattening the temperature gradient. For the case of $\alpha=0$ and $\delta=3/2$, the loop-top temperature has lowered to a warm coronal value of 2.78~MK, much more physically reasonable than that (15.2~MK) for $\alpha$=0 and $\delta=0$ (Figures \ref{fig04}a and b). Meanwhile, the dominance of turbulence in the corona ensures that a coronal depletion (or TR excess) of DEM still survives in all cases (Figure \ref{fig04}c).

The physical reasons for the reduction of heat flux can be understood as follows. First, the case of footpoint heating means a redistribution of energy deposition, which removes an amount of heating from the upper part of a loop and adds it to the lower part. Since the balance of energy in the corona is primarily between heating and conductive losses (see Figure \ref{fig02}), the depletion of the coronal heating should be mainly compensated for by a depression of downward transporting heat flux \citep[][]{Klimchuk19}. Second, for a loop with nonuniform cross section, the term of thermal conduction $E_{\scriptscriptstyle\! C}$ can be expanded as
\begin{equation}
E_{\scriptscriptstyle\! C}=-\frac{dF_{\scriptscriptstyle\! C}}{ds}-F_{\scriptscriptstyle\! C}\frac{d\ln A}{ds},
\end{equation}
where the second term on the right hand side arises from the variation of  cross section. If the loop cross section expands with height ($d\ln A/ds>0$), the contraction of cross section inversely toward the loop base acts as a bottle neck to obstruct the downward transport of heat flux \citep{Antiochos76}, whose effect is similar to an additional local heating \citep{Vesecky79}. Since the heat flux typically reaches its maximum near the loop base (see Figures \ref{fig01}d and \ref{fig04}d), the overall effect of the cross-sectional expansion is equivalent to a footpoint heating, which in turns reduces the heat flux from the corona.

\section{Numerical Modeling}\label{sec5}

\begin{figure}
\epsscale{1}
\plotone{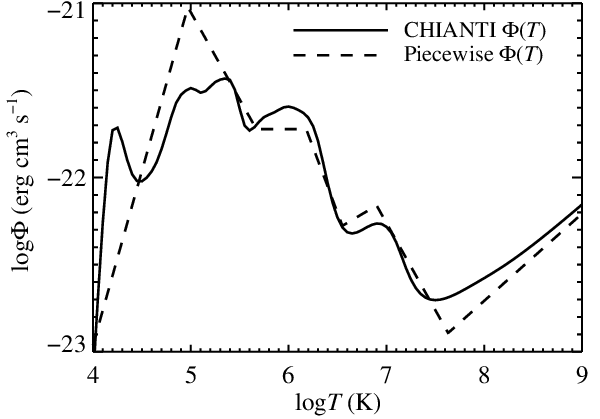}
\caption{Radiative loss functions used for the numerical loop modeling. The solid line represents the radiation function computed with the CHIANTI atomic database, and the dashed line denotes a piecewise power-law function that serves as an approximation of the realistic CHIANTI function. 
\label{fig05}}
\end{figure}

We further resort to numerical loop modeling for a check of the analytical modeling. Here we allow for a continuous transition between the importance of collisions and turbulence (rather than the clear-cut division between them as assumed in the analytical hybrid model), by which the mean free path pertinent to thermal conduction is formulated as
\begin{equation}
\lambda=\frac{\lambda_C\lambda_T}{\lambda_C+\lambda_T}.
\end{equation}
Moreover, the numerical modeling incorporates some other physics inapplicable for the analytical modeling, including gravitational stratification as well as improved functions of radiation, heating, and cross-sectional area.

We adopt a piecewise power-law function formulated in \citet[][]{Klimchuk08} as the radiative loss function for modeling. As shown in Figure \ref{fig05}, this formalized function serves as a quite good approximation of the ``realistic" radiation function computed with the CHIANTI atomic database \citep{Dere97,delZanna21}. It is noted that the piecewise function has been used in a variety of numerical approaches \citep[e.g.,][]{Klimchuk08,Cargill12,Bradshaw13,Barnes16}. 

To parametrize a nonuniform footpoint heating, we use an exponential function 
\begin{equation}
E_{\scriptscriptstyle\! H}(s)=H_b\exp\left(-\frac{s}{s_{\scriptscriptstyle\! H}}\right),
\end{equation}
where $H_b$ is the heating rate at the loop base and $s_{\scriptscriptstyle\! H}$ is the heating scale height. Compared with the explicitly temperature-dependent heating function used in the analytical modeling, the current distance-dependent function seems more appropriate in that the heating profile does not alter passively with the change of temperature structure, hence being more commonly used in numerical modeling \citep[e.g.,][]{Serio81,Aschwanden02,Ni22}. As pointed out above, a heating too concentrated near loop base (equivalent to a too small value of $s_{\scriptscriptstyle\! H}$) may lead to TNE inside a loop \citep{Klimchuk10,Froment18,Klimchuk19}. Even without TNE, the loop under such heating might be on the verge of thermal instability \citep[TI,][]{Parker53,Winebarger03,Klimchuk19b}. According to the rule of thumb made by \citet{Klimchuk19}, we choose a heating scale height of 0.5$L$ for the case of footpoint heating, which yields a high enough degree of heating stratification but still guarantees a thermally stable equilibrium of the loop.

As for loop cross-sectional expansion, we also prefer an explicitly distance-dependent areal function for numerical modeling, although the adoption of temperature-dependent ones makes the model possibly amenable to analytical solutions \citep[e.g.,][]{Levine80,Rabin91,Martens10}. We adapt the areal function used in \citet{Cargill22} and rewrite it into
\begin{equation}
A(s)=A_b\left[1+(\Gamma-1)\sin^2\left(\frac{\pi \min(s,s_{\scriptscriptstyle\! A})}{2s_{\scriptscriptstyle\! A}}\right)\right],
\end{equation}
where $A_b$ is the cross-sectional area at the loop base, $\Gamma$ ($\Gamma\ge1$) is the expansion factor defined as the ratio of apex to base cross-sectional areas, and $s_{\scriptscriptstyle\! A}$ ($0<s_{\scriptscriptstyle\! A}\le L$) is a scale length localizing the cross-sectional expansion. For large values of $\Gamma$ and $s_{\scriptscriptstyle\! A}=L$, the above areal function approximately depicts the cross-sectional variation of a magnetic flux tube controlled by a line dipole \citep{Antiochos80}. Although the rapid decay of coronal magnetic fields with height as revealed from magnetic modeling implies a large cross-sectional expansion of coronal loops \citep[e.g.,][]{Asgari12,Chen22}, observations typically demonstrate a less-than-expected expansion for the loops: they either keep a roughly constant cross section or just exhibit a modest cross-sectional expansion \citep[e.g.,][]{Klimchuk00,Klimchuk20,Williams21}. To reconcile with the observations, we set a moderate expansion factor of $\Gamma=5$ (with the loop width varying within a factor of $\sqrt{\Gamma}\approx2.2$) and a short expansion scale length of $s_{\scriptscriptstyle\! A}=0.2L$ (localized expansion near the loop base). For comparison, we also model the case of $s_{\scriptscriptstyle\! A}=0.8L$ whose cross-sectional expansion extends over most portion of the loop.

We consider a gravitationally stratified loop with a half-length of 30~Mm as specified before. At the footpoint of the loop, we set a pressure of $P_b=0.55$ dyne cm$^{-2}$ and temperature of $T_b=2\times10^4$~K \citep[cf,][]{Klimchuk88,Bradshaw10}, which yield a loop-base density of $1\times10^{11}$~cm$^{-3}$. Starting from the specified pressure and temperature as well as vanishing heat flux at the bottom boundary, we implement a shooting method to numerically solve the one-dimensional hydrostatic equations that involve the variations of pressure, temperature, and heat flux along the loop. We iteratively adjust the value of loop-base heating rate ($H_b$) and integrate the equations until the resultant heat flux vanishes at the loop apex too, from which other loop-top properties such as $T_a$ and $H_a$ are returned at the same time. The numerical integration incorporates a fourth-order Runge--Kutta calculator and an adaptive mesh refinement (with a refinement level up to 16) to improve the accuracy of the solutions.

\begin{figure*}
\epsscale{0.85}
\plotone{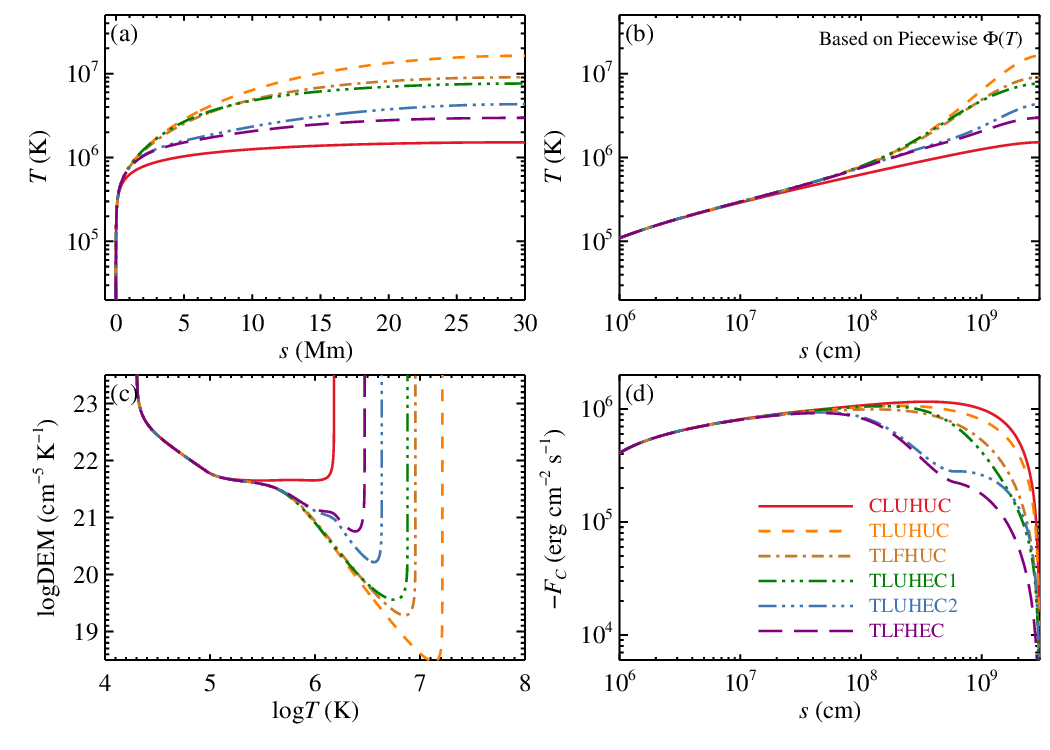}
\caption{Numerical solutions of a gravitationally stratified coronal loop computed for different parameters (discriminated with different colors and line styles), with the panels showing profiles of loop temperature (a and b), DEM (c), and heat flux (d), respectively. The detailed model parameters are listed in Table \ref{tb2}. Note that the results are derived using the piecewise power-law radiative loss function.
\label{fig06}}
\end{figure*}

\begin{deluxetable*}{ccccccchcc}[t]
\tablecaption{Representative Input Parameters and Output Results of the Numerical Modeling}\label{tb2}
\tablehead{
\colhead{Model} & \multicolumn{6}{c}{Input Parameters} & \colhead{} & \multicolumn{2}{c}{Output Results}\\
\cline{2-7} \cline{9-10}
\colhead{Name\tablenotemark{\scriptsize a}} & \colhead{$L$} & \colhead{$P_b$} & \colhead{$\lambda_T$} & \colhead{$s_{\scriptscriptstyle\! H}$} & \colhead{$\Gamma$} & \colhead{$s_{\scriptscriptstyle\! A}$} & \colhead{} & \colhead{$T_a$\tablenotemark{\scriptsize b}} & \colhead{$H_a$\tablenotemark{\scriptsize b}}\\
\colhead{} & \colhead{\footnotesize (Mm)} & \colhead{\footnotesize (dyne cm$^{-2}$)}  & \colhead{\footnotesize (km)} & \colhead{\footnotesize ($L$)} & \colhead{} & \colhead{\footnotesize ($L$)} & \colhead{} & \colhead{\footnotesize (MK)} & \colhead{\footnotesize ($\times$10$^{-4}$ erg cm$^{-3}$ s$^{-1}$})}
\startdata
CLUHUC & 30 & 0.55 & $\infty$ & $\infty$ & 1 & \nodata & & 1.52 (1.51) & 6.89 (7.38)\\
TLUHUC & 30 & 0.55 & 10 & $\infty$ & 1 & \nodata & & 16.3 (15.0) & 4.00 (3.80)\\
TLFHUC & 30 & 0.55 & 10 & 0.5 & 1 & \nodata & & 9.04 (8.38)& 1.28 (1.22)\\
TLUHEC1 & 30 & 0.55 & 10 & $\infty$ & 5 & 0.8 & & 7.61 (7.04)& 1.23 (1.17)\\
TLUHEC2 & 30 & 0.55 & 10 & $\infty$ & 5 & 0.2 & & 4.32 (4.62)& 1.45 (1.59)\\
TLFHEC & 30 & 0.55 & 10 & 0.5 & 5 & 0.2 & & 2.97 (3.17) & 0.568 (0.628)
\enddata
\tablenotetext{\scriptsize a}{\footnotesize Description of the abbreviations in model names: CL is for wholly collisional loop, TL for strongly turbulent loop, UH for uniform heating, FH for footpoint-concentrated heating, UC for uniform loop cross section, and EC for expanding cross section with height.}
\tablenotetext{\scriptsize b}{\footnotesize Values outside (inside) parentheses denote results derived using the piecewise power-law (CHIANTI) radiative loss function.}
\end{deluxetable*}

Figure \ref{fig06} displays numerical solutions of the loop computed for different model parameters, with the panels organized the same as in Figure \ref{fig04}. Some representative input parameters and output results for each model case are listed in Table \ref{tb2}. As shown in the figure, the numerical solutions are in general coincidence with the patterns already revealed in the analytical modeling, except that the numerically derived quantities vary more smoothly along the loop, a result of the treatment of a smooth transition of $\lambda$. In comparison to the wholly collision case with uniform heating and cross section (CLUHUC), the inclusion of only strong turbulence (with a $\lambda_T$ of 10~km, TLUHUC) elevates the loop-top temperature by even an order of magnitude (Figures \ref{fig06}a and b), but just marginally affects the heat flux (Figure \ref{fig06}d). Nevertheless, both a footpoint heating (TLFHUC) and a cross-sectional expansion (TLUHEC) can effectively lower down the unrealistically high coronal temperature  by notably reducing the heat flux in the turbulent corona. The effect is more prominent for footpoint-localized cross-sectional expansion (TLUHEC2) than for global expansion (TLUHEC1), because the former one  corresponds to a higher relative expansion rate ($d\ln A/ds$) near the loop base where the heat flux peaks. An amalgamation of the footpoint heating and localized cross-sectional expansion (TLFHEC) eventually reduces the loop-top temperature to $\sim$3~MK, in good agreement with the values  observed in AR core loops \citep{delZanna18}. Finally, all the turbulent cases produce a depletion of the DEM in the corona (Figure \ref{fig06}c), also consistent with the observations. 

\section{Discussion and Conclusions}\label{sec6}
In this work we generalize the hybrid loop mode of \citet{Emslie22} by incorporating nonuniform heating and cross section that are both characterized by a power-law function of temperature. Based on the hybrid model solutions, we construct scaling laws for various situations. With the inclusion of a turbulence-dominated part, an accurate solution of the analytical hybrid model actually involves a numerical evaluation of the finite integral $I(\theta_0)$ (Equation (\ref{Itheta0})). At the TDC limit, nevertheless, our scaling laws (Equations (\ref{sltat1}--\ref{slhat1}), (\ref{sltat2}--\ref{slhat2}), and (\ref{sltat3}--\ref{slhat3})) give a robust estimate of loop-top properties computed fully with the analytical model but use a more simple and straightforward approach (see Figure~\ref{fig03}).

Compared with the ``unified" scaling laws at the CDC limit (Equations (\ref{sltac}) and (\ref{slhac})), the TDC scaling laws show substantial changes. First, the loop-top properties are additionally power-law functions of the turbulent mean free path, with $T_a$ ($H_a$) increasing (decreasing) with the decrease of $\lambda_T$. Second and more importantly, their functional forms vary from situation to situation, whose classification depends on the specification of the heating and/or areal parameters. It is found that both a sufficiently footpoint-concentrated heating and a cross-sectional expansion with height can effectively weaken (strengthen) the negative (positive) power-law dependence of $T_a$ ($H_a$) on $\lambda_T$ (see Table \ref{tb1} and Figure \ref{fig03}). In this sense, the commonly modeled case of uniform heating and cross section, which possesses the strongest functional dependence between $T_a$ and $\lambda_T$, may not adequately describe the characteristics of a turbulent coronal loop.

Such new patterns revealed by the generalized TDC scaling laws arise from a much weaker (or even inverse) dependence of the conduction coefficient on temperature ($\kappa\propto T^{-1/2}$) that is introduced by turbulent scattering. Since heat flux is the product of  conduction coefficient and temperature gradient, the suppression of the conduction coefficient needs to be compensated for by a sharpening of the temperature gradient if the heat flux remains relatively unchanged in the turbulence-dominated part, as is the case of uniform heating and cross section (see Figure \ref{fig01}). On the other hand, in case of footpoint heating and/or cross-sectional expansion, the heat flux transporting downward from the corona is notably reduced. If the region of reduced heat flux overlaps with the turbulence-dominated part, the reduction of the heat flux must predominately rely on a backward mitigation of the temperature gradient (see Figure (\ref{fig04})). Through numerical modeling that incorporates more realistic conditions, the above physical picture is further consolidated (see Figures (\ref{fig06}) and (\ref{fig08})).

In fact, the reduction of heat flux by footpoint heating and/or cross-sectional expansion takes effect regardless of collisional or turbulent situations \citep{Klimchuk19,Cargill22}. In case of a wholly collision-dominated loop, a strong dependence of the conduction coefficient on temperature ($\kappa\propto T^{5/2}$) implies that even a small adjustment of the temperature can considerably alter the conduction coefficient, and consequently the heat flux. Therefore, a notable reduction of the heat flux in the corona just requires a slight decrease of the coronal temperature, as revealed in Figure \ref{fig03}a (loop-top temperatures at the CDC limit for various heating and/or areal parameters). In this sense, the effects of footpoint heating and cross-sectional expansion become more and more prominent as the turbulence strength increases.

These results have important implication for loops in ARs, where magnetic fields are locally concentrated. On one hand, the concentration of magnetic fields facilitates the occurrence of strong enough turbulence in AR loops. Such level of turbulence naturally results in a DEM profile rising toward low temperatures, more consistent with observations than that predicted by the wholly collisional model \citep{Gontikakis23}. Without the need of other complex mechanisms, this simple turbulent scenario seems rather compelling \citep{Emslie22}. On the other hand, the decay of magnetic fields with height implies a decrease of background heating rate as well as an increase of cross-sectional area toward the loop apex, both of which effectively lower the sensitivity of loop temperature to turbulence strength. Therefore, such AR loops can bear relatively stronger turbulence while still keeping a physically reasonable temperature for nonflaring loops.

It should be pointed out that our current modeling, both analytical and numerical, is restricted in hydrostatic loops. \citet{Bradshaw20} have established scaling laws for dynamic loops of uniform heating and cross section, finding that the scaling laws for low Mach number flows just differ slightly from the static case. In a next step work, we plan to generalize the dynamic loop scaling laws by also incorporating nonuniform heating and cross section, whose effects will be further investigated in terms of the Mach number of flows.

Even for highly dynamic loops (e.g., loops powered by impulsive flare or nanoflare energy release) that undergo heating--cooling cycles, an expansion of the loop cross section with height should be still of physical significance. In the first-stage conductive cooling, the cross-sectional expansion suppresses the downward transport of heat flux and dilutes the evaporative flow to the corona \citep{Reep22}, while in the second-stage radiative cooling, it prevents the evaporated materials from draining back to the chromosphere until a catastrophic cooling occurs \citep{Cargill22,Reep23}. Both processes (that cause the reduction of both conductive and radiative losses) can account for a longer-than-expected cooling time observed in some solar flares \citep{Ryan13} without the need of additional heating. Nevertheless, an in-depth investigation of such processes is beyond the scope of this work.

Software procedures called to generate the analytical and numerical solutions in this work are available from the authors on request. The routines are coded with Interactive Data Language (IDL), and we encourage users to run them under the Solar Software \citep[SSW,][] {Freeland98} environment.\\
	
This work was supported by National Natural Science Foundation of China under grants 11733003 and 12127901. Y.D. is also sponsored by National Key R\&D Program of China under grants 2019YFA0706601 and 2020YFC2201201, as well as Frontier Scientific Research Program of Deep Space Exploration Laboratory under grant 2022--QYKYJH--HXYF--015.

\appendix
\section{TDC Scaling Laws in Terms of the Turbulent Knudsen Number}
Using Equation (\ref{lambtc}), the TDC scaling laws presented in Section \ref{sec32} can be readily rewritten in terms of the turbulent Knudsen number \citep[$\lambda_T/L$,][]{Bradshaw20}, a parameter case independent. The adaptation does not change the power-law dependence of the loop-top properties on $\lambda_T$.

For situation T1 ($\mu_2+1<\nu_2+1<0$),
\begin{equation}
\begin{aligned}
T_a& \approx\left(\frac{\phi_1\chi_0}{k_{\scriptscriptstyle\! B}^2\kappa_{\scriptscriptstyle 0C}}\right)^{2/(1-2\alpha)}\left(PL\right)^{2(1-\gamma-\alpha)/3(1-2\alpha)}\\
&\ \times\left(\frac{1}{ c_{\scriptscriptstyle\! R}}\frac{\lambda_T}{L}\right)^{-2(5+\gamma+\alpha)/3(1-2\alpha)}\ \ \ [\mathrm{T1}]
\end{aligned}
\end{equation}
and
\begin{equation}
\begin{aligned}
H_a&\approx\frac{(7+2\alpha+4\delta)(1+2\alpha+4\delta)}{4(3-2\gamma+4\delta)(-3-2\gamma+4\delta)\phi_1}\left(\frac{\phi_1\chi_0}{k_{\scriptscriptstyle\! B}^2\kappa_{\scriptscriptstyle 0C}}\right)^{1/(1-2\alpha)}\\
&\ \times P^{(4-\gamma-7\alpha)/3(1-2\alpha)}L^{-(2+\gamma-5\alpha)/3(1-2\alpha)}\\
&\ \times \kappa_{\scriptscriptstyle 0C}\left(\frac{1}{ c_{\scriptscriptstyle\! R}}\frac{\lambda_T}{L}\right)^{-(2+\gamma+7\alpha)/3(1-2\alpha)}.\ \ \ [\mathrm{T1}]
\end{aligned}
\end{equation}

For situation T2 ($\mu_2+1<0<\nu_2+1$),
\begin{equation}
\begin{aligned}
T_a&\approx\left(\frac{\phi_2\chi_0}{k_{\scriptscriptstyle\! B}^2\kappa_{\scriptscriptstyle 0C}}\right)^{1/(1+2\delta)}\left(PL\right)^{(3-2\gamma+4\delta)/6(1+2\delta)}\\
&\ \times\left(\frac{1}{ c_{\scriptscriptstyle\! R}}\frac{\lambda_T}{L}\right)^{-(9+2\gamma-4\delta)/6(1+2\delta)}\ \ \ [\mathrm{T2}] 
\end{aligned}
\end{equation}
and
\begin{equation}
\begin{aligned}
H_a&\approx\frac{3(1+2\alpha+4\delta)}{2(3-2\gamma+4\delta)(3+2\gamma-4\delta)\phi_2}\left(\frac{\phi_2\chi_0}{k_{\scriptscriptstyle\! B}^2\kappa_{\scriptscriptstyle 0C}}\right)^{1/2(1+2\delta)}\\
&\ \times P^{(15-2\gamma+28\delta)/12(1+2\delta)}L^{-(9+2\gamma+20\delta)/12(1+2\delta)}\\
&\ \times \kappa_{\scriptscriptstyle 0C}\left(\frac{1}{ c_{\scriptscriptstyle\! R}}\frac{\lambda_T}{L}\right)^{(3-2\gamma+28\delta)/12(1+2\delta)}.\ \ \ [\mathrm{T2}]
\end{aligned}
\end{equation}

For situation T3 ($0<\mu_2+1<\nu_2+1$),
\begin{equation}
\begin{aligned}
T_a&\approx\left(\frac{\phi_3\chi_0}{k_{\scriptscriptstyle\! B}^2\kappa_{\scriptscriptstyle 0C}}\right)^{2/(5+2\gamma)}\left(PL\right)^{2/(5+2\gamma)}\\
&\ \times\left(\frac{1}{ c_{\scriptscriptstyle\! R}}\frac{\lambda_T}{L}\right)^{-2/(5+2\gamma)}\ \ \ [\mathrm{T3}]
\end{aligned}
\end{equation}
and
\begin{equation}
\begin{aligned}
H_a&\approx\frac{1+2\alpha+4\delta}{4(-3-2\gamma+4\delta)\phi_3}\left(\frac{\phi_3\chi_0}{k_{\scriptscriptstyle\! B}^2\kappa_{\scriptscriptstyle 0C}}\right)^{1/(5+2\gamma)}\\
&\ \times P^{2(3+\gamma)/(5+2\gamma)}L^{-2(2+\gamma)/(5+2\gamma)}\\
&\ \times\kappa_{\scriptscriptstyle 0C}\left(\frac{1}{ c_{\scriptscriptstyle\! R}}\frac{\lambda_T}{L}\right)^{2(2+\gamma)/(5+2\gamma)}.\ \ \ [\mathrm{T3}]
\end{aligned}
\end{equation}

\section{Special Turbulent Situations of $\nu_2+1=0$ and $\mu_2+1=0$}
\begin{figure}
\epsscale{0.9}
\plotone{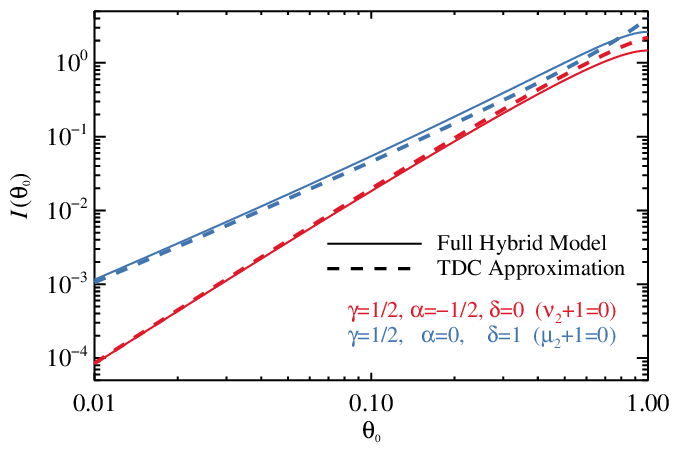}
\caption{Variations of $I(\theta_0)$ vs.\ $\theta_0$ for two combinations of $\gamma$, $\alpha$, and $\delta$ values. The case of $\gamma=1/2$, $\alpha=-1/2$, and $\delta=0$ (red) belongs to the situation of $\nu_2+1=0$, whereas the case of $\gamma=1/2$, $\alpha=0$, and $\delta=1$ (blue) is for the situation of $\mu_2+1=0$. The integral $I(\theta_0)$ is evaluated with the full hybrid model (solid) and the TDC approximation (dashed), respectively.  
\label{fig07}}
\end{figure}

\begin{figure*}
\epsscale{0.83}
\plotone{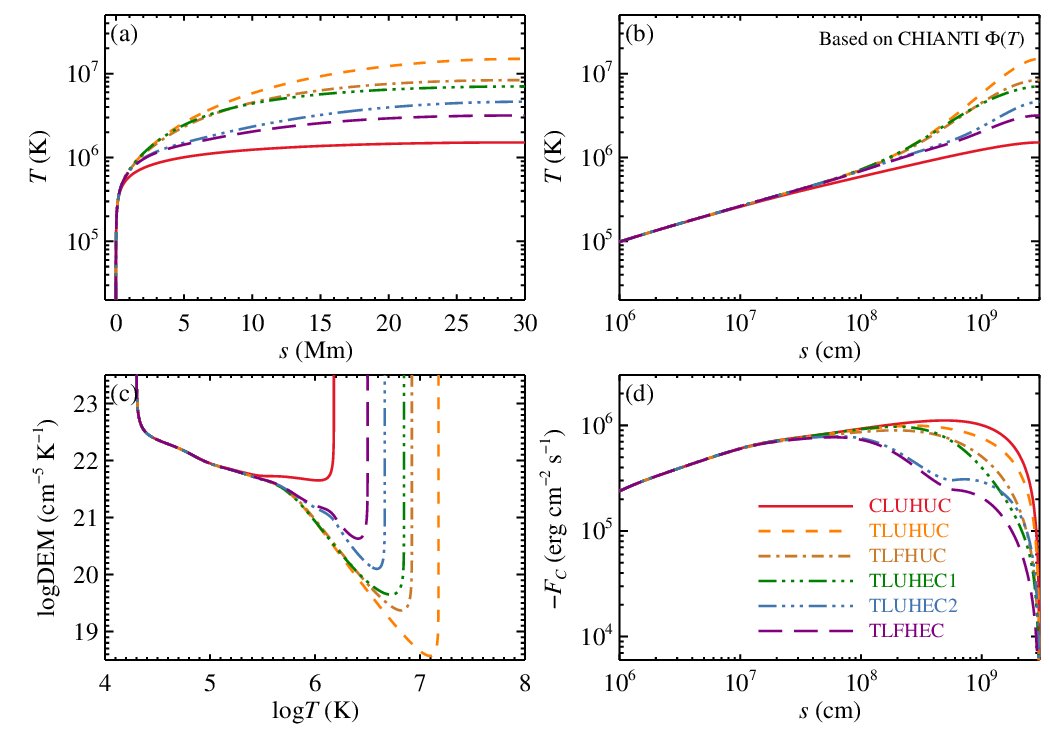}
\caption{The same as Figure \ref{fig06}, except that the solutions are derived using the CHIANTI radiative loss function.
\label{fig08}}
\end{figure*}

In addition to the three turbulent situations T1--T3, there are two other special situations: $\mu_2+1<\nu_2+1=0$ and $0=\mu_2+1<\nu_2+1$, respectively. For the situation of $\nu_2+1=0$, the original form of the turbulent first integral (Equation (\ref{fiturb})) needs to be modified to incorporate a logarithmic function of $\zeta$. With this modification, the expressions of $\xi$ and $I_2(\theta_0)$ now become
\begin{equation}
\xi=\frac{3\left[3-(1-\gamma-\alpha)\theta_0^{2+\gamma+\alpha}\right]\theta_0^{-(2+\gamma+\alpha)}}{(1-\gamma-\alpha)(2+\gamma+\alpha)(1-3\ln\theta_0)}\label{xi12}
\end{equation}
and
\begin{equation}
\begin{aligned}
I_2(\theta_0)=\int_{\zeta_0}^1\left(-\frac{1-\zeta^{\mu_2+1}}{\mu_2+1}-\xi\ln\zeta
\right)^{-1/2}d\zeta.
\end{aligned}
\end{equation}
At the TDC limit ($\theta_0\to0$), the parameter $\xi$ is approximated as
\begin{equation}
\xi\approx\frac{9\theta_0^{-(2+\gamma+\alpha)}}{(1-\gamma-\alpha)(2+\gamma+\alpha)(1-3\ln\theta_0)},
\end{equation}
and the finite integral $I(\theta_0)$ (from Equation (\ref{Itheta0t})) accordingly reduces to
\begin{equation}
\begin{aligned}
I(\theta_0)&\approx\left(\frac{7+2\delta}{1+2\delta}\right)^{1/2}\frac{\theta_0^{3/2}}{\xi^{1/2}}\int_0^1(-\ln\zeta)^{-1/2}d\zeta\\
&=\left[\frac{(7+2\delta)\pi}{1+2\delta}\right]^{1/2}\frac{\theta_0^{3/2}}{\xi^{1/2}}\\
&\approx\left[\frac{(7+2\delta)(1-\gamma-\alpha)(2+\gamma+\alpha)\pi}{9(1+2\delta)}\right]^{1/2}\\
&\ \times(1-3\ln\theta_0)^{1/2}\theta_0^{(5+\gamma+\alpha)/2}.\label{Itheta012}
\end{aligned}
\end{equation}
	
For the situation of $\mu_2+1=0$,  the expressions of $\xi$ and $I_2(\theta_0)$ are similarly modified to
\begin{equation}
\xi=\frac{(2+\gamma+\alpha)(5+\gamma+\alpha)(1-3\ln\theta_0)}{3\left[(5+\gamma+\alpha)-3\theta_0^{2+\gamma+\alpha}\right]}
\end{equation}
and
\begin{equation}
\begin{aligned}
I_2(\theta_0)=\int_{\zeta_0}^1\left[\ln\zeta+\frac{\xi(1-\zeta^{\nu_2+1})}{\nu_2+1}\right]^{-1/2}d\zeta.
\end{aligned}
\end{equation}
At the TDC limit, the approximations of $\xi$ and $I(\theta_0)$ consequently lead to
\begin{equation}
\xi\approx\frac{2+\gamma+\alpha}{3}(1-3\ln\theta_0)
\end{equation}	
and
\begin{equation}
\begin{aligned}
I(\theta_0)&\approx\left(\frac{7+2\delta}{1+2\delta}\right)^{1/2}\frac{\theta_0^{3/2}}{\xi^{1/2}}\int_0^1\left(\frac{1-\zeta^{\nu_2+1}}{\nu_2+1}\right)^{-1/2}d\zeta\\
&=\left[\frac{7+2\delta}{(1+2\delta)(\nu_2+1)}\right]^{1/2}B\left(\frac{1}{\nu_2+1},\frac{1}{2}\right)\frac{\theta_0^{3/2}}{\xi^{1/2}}\\
&\approx\left[\frac{3(7+2\delta)}{2(2+\gamma+\alpha)^2}\right]^{1/2}B\left(\frac{1+2\delta}{2(2+\gamma+\alpha)},\frac{1}{2}\right) \\
&\  \times\frac{\theta_0^{3/2}}{(1-3\ln\theta_0)^{1/2}}.\label{Itheta023}
\end{aligned}
\end{equation}

The case of $\nu_2+1=0$ connects situations T1 and T2, whereas the case of $\mu_2+1=0$ is the transition point between situations T2 and T3. Figure \ref{fig07} plots the variations of $I(\theta_0)$ versus $\theta_0$ for two combinations of $\gamma$, $\alpha$, and $\delta$ values, which belong to the situations of $\nu_2+1=0$ and $\mu_2+1=0$, respectively. The integral $I(\theta_0)$ is evaluated with the full hybrid model (from Equation (\ref{Itheta0})) and the TDC approximation (Equation (\ref{Itheta012}) or (\ref{Itheta023})), respectively. For small values of $\theta_0$, the simple TDC approximation gives an excellent estimation of the finite integral $I(\theta_0)$. However, the TDC approximation now involves an additional term of $1-3\ln\theta_0$, which makes $\theta_0$ transcendental functions of $\lambda_T$ (see Equation (\ref{theta0lambt})). Therefore, the construction of scaling laws in power-law functions of $\lambda_T$, as done before, becomes unpractical for both special situations.

\section{Numerical Solutions Computed with the CHIANTI Radiative Loss Function}
In the numerical modeling, we also construct loop solutions using the ``realistic" CHIANTI radiative loss function (plotted as the solid line in Figure \ref{fig05}). The radiation function is calculated under an assumption of coronal elemental abundance \citep[e.g.,][]{Schmelz12}, where the abundances of elements with low first ionization potential (FIP) exhibit an enrichment over the photospheric values \citep{Meyer85,Feldman92}. Figure \ref{fig08} displays the numerical solutions computed with the CHIANTI function. Compared to Figure \ref{fig06}, it is seen that the CHIANTI-based solutions are quantitatively in good agreement with those derived using the piecewise radiation function. Taking the loop-top properties ($T_a$ and $H_a$) as a benchmark, the deviations between the two sets of solutions are typically less than 10\% (see Table \ref{tb2}). This consistency validates the use of the  piecewise power-law function in common numerical modeling. 


\end{document}